\documentclass[letterpaper,english,reprint,aps,longbibliography]{revtex4-2}
\usepackage[T1]{fontenc}
\usepackage[utf8]{inputenc}
\usepackage{color}
\usepackage{babel}
\usepackage{amsmath}
\usepackage{amssymb}
\usepackage{graphicx}
\usepackage[pdfusetitle,
 bookmarks=true,bookmarksnumbered=false,bookmarksopen=false,
 breaklinks=true,pdfborder={0 0 1},backref=false,colorlinks=true]
 {hyperref}



\begin{document}
\title{Black-Hole Signatures in the Finite-Temperature Critical Ising Chain}
\author{Zuo Wang}
\affiliation{Institute for Theoretical Physics, School of Physics, South China Normal University, Guangzhou 510006, China}
\affiliation{Key Laboratory of Atomic and Subatomic Structure and Quantum Control (Ministry of Education), Guangdong Basic Research Center of Excellence for Structure and Fundamental Interactions of Matter, School of Physics, South China Normal University, Guangzhou 510006, China}
\affiliation{Guangdong Provincial Key Laboratory of Quantum Engineering and Quantum Materials, Guangdong-Hong Kong Joint Laboratory of Quantum Matter, South China Normal University, Guangzhou 510006, China}
\author{Liang He}
\email{liang.he@scnu.edu.cn}

\affiliation{Institute for Theoretical Physics, School of Physics, South China Normal University, Guangzhou 510006, China}
\affiliation{Key Laboratory of Atomic and Subatomic Structure and Quantum Control (Ministry of Education), Guangdong Basic Research Center of Excellence for Structure and Fundamental Interactions of Matter, School of Physics, South China Normal University, Guangzhou 510006, China}
\affiliation{Guangdong Provincial Key Laboratory of Quantum Engineering and Quantum Materials, Guangdong-Hong Kong Joint Laboratory of Quantum Matter, South China Normal University, Guangzhou 510006, China}
\begin{abstract}
We demonstrate that the finite-temperature critical transverse-field Ising chain exhibits quantitative signatures of black-hole physics in its dual gravitational description within the AdS/CFT correspondence. Its finite-temperature dynamics and thermodynamics are consistently captured by a mixed thermal-AdS/BTZ black hole saddle, leading to three mutually compatible observations. First, antipodal excitation transport collapses onto a universal temperature-dependent curve determined by the relative AdS and BTZ contributions to the gravitational partition function, reflecting horizon absorption. Second, in the high-temperature regime, the retarded response exhibits exponential relaxation governed by the lowest quasi-normal mode of the dual black hole. Third, the temperature derivative of the von Neumann entropy develops a pronounced minimum at a temperature consistent with the Hawking--Page transition. These results identify critical quantum spin chains as minimal and experimentally accessible platforms for probing dynamical and thermodynamic aspects of quantum black holes in controllable many-body systems. 
\end{abstract}
\maketitle
\emph{Introduction}.---The AdS/CFT correspondence posits a deep equivalence between quantum gravity in asymptotically anti--de Sitter (AdS) spacetimes and conformal field theories (CFTs) defined on their boundaries, providing a concrete and nonperturbative framework in which gravitational phenomena can be studied using many-body quantum physics \cite{Maldacena1999IJTP,Gubser_PLB_1998,Witten1998ATMP}. Beyond its conceptual significance, this duality has reframed gravity in operational terms---relating horizons, black holes, and spacetime dynamics to correlation functions, transport, and relaxation in quantum systems---and has motivated the search for laboratory platforms capable of emulating quantum-gravitational effects. A striking realization of this program is provided by the nearly $\text{AdS}_{2}/\text{CFT}_{1}$ correspondence in Sachdev--Ye--Kitaev--type models, where quantum simulators have accessed wormhole dynamics and horizon-related phenomena \cite{Sachdev_Ye_1993_PRL,Kitaev_2015_talk,Maldacena_Stanford_2016_PRD,Jafferis_2022_nature,Susskind_Swingle_2020_arxiv}. A natural next frontier is to extend such table-top probes to two-dimensional CFTs, where higher-dimensional black-hole physics, including thermodynamics and dissipative dynamics, becomes accessible.

Among candidate theories, the Ising CFT occupies a unique position. As the minimal nontrivial CFT, it has been argued to admit a three-dimensional gravitational dual that supports black-hole states, even though no simple semiclassical geometric description is available \cite{Castro_2012_PRD,Jian_2020_JHEP}. Importantly, the Ising CFT is not merely a theoretical construct: it emerges as the universal low-energy description of critical one-dimensional quantum spin chains \cite{Sachdev_1999_book_QPT}, enabling a direct confrontation between holographic ideas and microscopic many-body physics. These critical spin chains can be engineered with high precision in both programmable analog simulators and digital quantum computing platforms, making them among the most experimentally accessible realizations of critical many-body dynamics and natural candidates for exploring holographic phenomena in controllable quantum systems \cite{Monroe2021RMP,haghshenas_PRL_2024,Bernien_Nature_2017,Keesling_Nature_2019,Ebadi2021nature,Scholl2021nature,Bluvstein_2022_nature,fang2024arxiv,kim2024arXiv}. From the gravitational perspective, finite temperature plays a decisive role: thermal states of the CFT correspond to bulk geometries that interpolate between thermal AdS and black-hole spacetimes \cite{Hawking_Page_1983_CMP,Witten_1998_ads_confinement}. This structure predicts a set of sharp, universal signatures---horizon absorption governing excitation transport, quasi-normal modes controlling late-time relaxation, and the Hawking--Page transition separating distinct thermal saddles---that should manifest coherently in boundary observables \cite{Birmingham_2002_PRL,Hawking_Page_1983_CMP,Konoplya_2011_RMP,Witten_1998_ads_confinement}.

\begin{figure}
\includegraphics{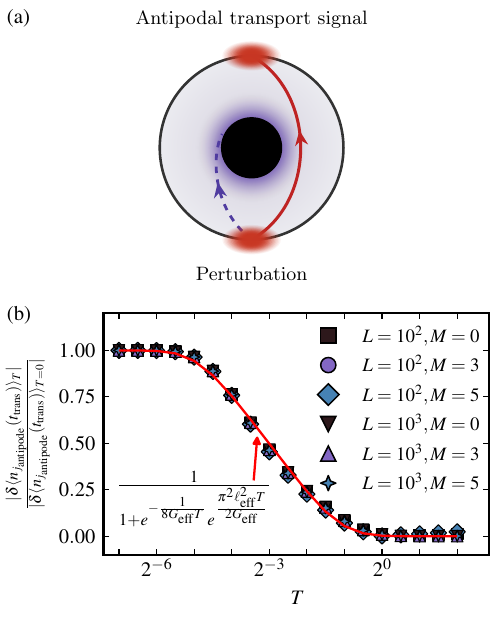}

\caption{(a) Schematic illustration of the finite-temperature critical spin chain and its dual gravitational description, comprising thermal AdS spacetime and a BTZ black hole. A localized perturbation generates excitations that correspond to particles propagating from the boundary into the bulk. They either reach the antipodal point through the AdS geometry (red arrow) or fall into the black hole (dashed purple arrow). (b) Temperature dependence of antipodal excitation transport for various system sizes $L$ and angular momenta $M$. All numerical data collapse onto a universal curve determined by the relative partition weights $Z_{\mathrm{AdS}}(T)/Z_{\mathrm{grav}}(T)$, shown as the red line, with effective parameters $\ell_{\mathrm{eff}}=1.28$ and $G_{\mathrm{eff}}=1.33$. See text for details. }
\label{Fig_1}
\end{figure}

\begin{figure}
\includegraphics[width=3.2in]{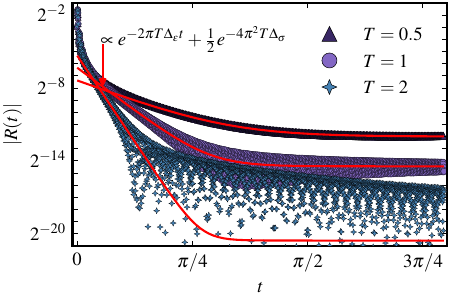}

\caption{Manifestation of the quasi-normal mode of the BTZ black hole in the critical spin chain. The time dependence of the spatially summed retarded response function $R(t)$ for the scalar perturbation in the high-temperature regime is computed for a critical spin chain of length $L=10^{3}$. The magnitude $|R(t)|$ exhibits a clear exponential decay governed by the lowest quasi-normal mode of the BTZ black hole, ($\exp(-2\pi T\Delta t)$ with $\Delta=1$), over an extended time window. Beyond the leading exponential behavior, $|R(t)|$ develops a temperature-dependent offset given by $\exp(-\pi^{2}T/2)/2$. See text for details.}

\label{Fig_2}
\end{figure}

\begin{figure}
\includegraphics[width=2.3in]{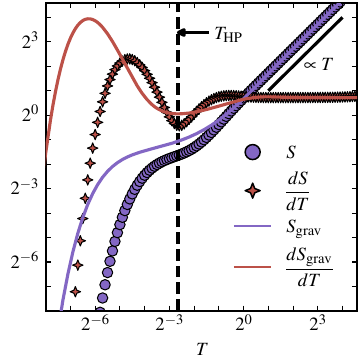}\caption{Entropy and its temperature derivative for the critical Ising spin chain and its dual gravitational system. The entropy of the spin chain, $S(T)$, is computed for system size $L=10^{3}$, while the entropy of the dual gravitational system, $S_{\mathrm{grav}}(T)$, is obtained directly from its partition function Eq.~(\ref{eq:Partition_function_gravity_system}). At low temperatures, the spin-chain entropy remains close to zero, consistent with the gravitational result $S_{\mathrm{grav}}\approx S_{\mathrm{AdS}}(T)=0$. At high temperatures, $S(T)$ exhibits a linear dependence on $T$, in agreement with the BTZ black hole--dominated scaling in the bulk description, $S_{\mathrm{grav}}(T)\approx S_{\mathrm{BTZ}}(T)\propto T$. In the intermediate temperature regime, $dS(T)/dT$ develops a pronounced minimum at $T=0.16\pm0.01$, in good agreement with the corresponding minimum of the gravitational entropy derivative, $dS_{\mathrm{grav}}(T)/dT$, which occurs at the Hawking--Page transition temperature $T_{\mathrm{HP}}=1/2\pi\approx0.16$. See text for details.}
\label{Fig_3}
\end{figure}

In this work, we address this question for the critical transverse-field Ising chain at finite temperature and demonstrate that it exhibits quantitative signatures of black-hole physics in its dual gravitational description within the AdS/CFT correspondence. Specifically, we find the following. (i) The finite-temperature excitation transport displays a universal behavior governed by the fundamental parameters of the dual gravitational system, directly reflecting the horizon devouring of the BTZ black hole (see Fig.~\ref{Fig_1}). (ii) In the high-temperature regime, the retarded response exhibits exponential decay governed by the characteristic quasi-normal mode of the dual black hole (see Fig.~\ref{Fig_2}). (iii) In the intermediate-temperature regime, the temperature derivative of the von Neumann entropy develops a pronounced local minimum at a temperature consistent with the Hawking--Page transition of the dual gravitational description (see Fig.~\ref{Fig_3}).

\emph{Finite-temperature critical spin chain and its gravitational dual.}---We consider the one-dimensional transverse-field Ising chain with Hamiltonian 
\begin{equation}
H=-\frac{L}{4\pi}\sum_{j=1}^{L}\left(Z_{j}Z_{j+1}+gX_{j}\right),\label{eq:TFIM}
\end{equation}
where $X_{j},Y_{j},$ and $Z_{j}$ are Pauli operators at site $j$, and $L$ is the system size. This model has a critical point at $g=1$ where its low-energy properties are governed by the Ising CFT \cite{Francesco_CFT_Book_2012}. The prefactor $L/(4\pi)$ ensures that the corresponding CFT is defined on a spatial circle of circumference $2\pi$ \cite{Bamba_2024_PRD,Sahay_2025_PRX}. We use units with $\hbar=k_{B}=1$.

According to the AdS/CFT correspondence \cite{Maldacena1999IJTP,Gubser_PLB_1998,Witten1998ATMP}, the critical-point physics of the system is expected to be related to gravitational physics in AdS spacetime. Indeed, at zero temperature $T=0$, recent studies \cite{Bamba_2024_PRD,Sahay_2025_PRX,imagawa2025arXiv,Wang_2025_arXiv} have shown that several critical spin chains exhibit intriguing properties consistent with a clear correspondence to gravitational physics in pure AdS spacetime. At finite temperature $T\neq0$, the AdS/CFT correspondence further predicts that a finite-temperature CFT corresponds to gravitational physics in AdS spacetime in the presence of a black hole \cite{Kurita_2005_PTP,Witten_2024_arxiv_introductionblackholethermodynamics}. Interestingly, recent studies of the Ising CFT have shown that its correlation functions in the high-temperature regime may reveal preliminary signatures of this correspondence \cite{Janik_2025_PRD}. This motivates us to systematically explore possible manifestations of black-hole physics in the critical spin chain at finite temperature.

To this end, we first construct a phenomenological dual gravitational description of the system via the AdS/CFT correspondence \cite{Maldacena1999IJTP,Gubser_PLB_1998,Witten1998ATMP} in order to identify the gravitational phenomena expected for the critical spin chain (\ref{eq:TFIM}) at finite temperature. The effective gravitational description consists of two contributions \cite{Nguyen_Brian_2018_JHEP,Miguel_2024_PRR}: thermal $(2+1)$-dimensional AdS spacetime and the $(2+1)$-dimensional Bañados--Teitelboim--Zanelli (BTZ) black hole \cite{Zanelli_1992_PRL,Zanelli_1993_PRD}. The partition function of the dual gravity system at temperature $T$ reads
\begin{equation}
Z_{\mathrm{grav}}(T)\simeq Z_{\mathrm{AdS}}(T)+Z_{\mathrm{BTZ}}(T),\label{eq:Partition_function_gravity_system}
\end{equation}
where $Z_{\mathrm{AdS}}(T)=e^{1/(8GT)}$ and $Z_{\mathrm{BTZ}}(T)=e^{T\pi^{2}\ell^{2}/(2G)}$ are the contributions from thermal AdS spacetime and the BTZ black hole, respectively. Here $\ell$ and $G$ denote the AdS radius and Newton’s constant.

From this effective description, we identify three characteristic manifestations of black-hole physics in the critical Ising chain. First, the presence of a $(2+1)$-dimensional BTZ black hole in the bulk implies that particles propagating from the boundary into the bulk eventually cross the horizon and disappear \cite{Sup_Mat} (see the schematic illustration in Fig.~\ref{Fig_1}). In the boundary theory, this process translates into the decay of the transport amplitude of excitations, which correspond to bulk particle propagation. As we demonstrate below, this simple horizon-absorption mechanism gives rise to a universal finite-temperature transport dynamics in the critical spin chain. Second, in the high-temperature regime where the BTZ black hole dominates the gravitational partition function, the characteristic fingerprint of the black hole---its quasi-normal modes \cite{Frolov_Black_hole_Physics_Book_1998,Konoplya_2011_RMP}---becomes manifest. On the boundary side, this implies that the decay of the response function is primarily governed by the quasi-normal modes of the dual black hole. Third, Eq.~(\ref{eq:Partition_function_gravity_system}) implies the existence of the Hawking--Page transition \cite{Hawking_Page_1983_CMP} at $T=T_{\mathrm{HP}}=1/(2\pi\ell)$, where thermal AdS spacetime and the BTZ black hole contribute equally to the partition function. At this temperature, the first derivative of the gravitational entropy with respect to temperature, $dS_{\mathrm{grav}}/dT$, exhibits a characteristic local minimum. Correspondingly, we show that the entropy of the critical spin chain displays the same thermodynamic signature. Together, these results demonstrate that several hallmark phenomena of black-hole physics emerge directly in the finite-temperature dynamics and thermodynamics of the critical spin chain.

\emph{Universal finite-temperature transport as a manifestation of matter absorption by the BTZ black hole.}---We begin with the most basic property of a black hole: matter crossing the horizon is irreversibly absorbed. For a generic black hole, a particle traveling from the boundary into the bulk is absorbed if its geodesic crosses the horizon and escapes otherwise. For the $(2+1)$-dimensional BTZ black hole, however, even null geodesics inevitably fall into the horizon due to the structure of the spacetime metric \cite{Sup_Mat}. Consequently, for antipodal transport---propagation from a boundary point to its antipode---the BTZ contribution never reaches the boundary regardless of the geodesic details {[}see Fig.~\ref{Fig_1}(a){]}. Let $N_{\mathrm{trans}}^{\mathrm{AdS}}$ denote the fraction of particles reaching the antipode in pure AdS spacetime. At finite temperature $T$, the transported fraction follows from the relative weights of the AdS and BTZ saddles, $N_{\mathrm{trans}}(T)=N_{\mathrm{trans}}^{\mathrm{AdS}}\cdot[Z_{\mathrm{AdS}}(T)/Z_{\mathrm{grav}}(T)]+0\cdot[Z_{\mathrm{BTZ}}(T)/Z_{\mathrm{grav}}(T)]$, which immediately gives $N_{\mathrm{trans}}(T)/N_{\mathrm{trans}}(0)=Z_{\mathrm{AdS}}(T)/Z_{\mathrm{grav}}(T)$. As we show below, this leads to a universal finite-temperature transport behavior in the critical spin chain determined solely by the parameters of the dual gravitational description. 

According to the AdS/CFT correspondence, boundary excitations corresponding to particles propagating into the bulk can be generated by a spatiotemporally localized perturbation \cite{Kinoshita_2023_JHEP}. For the critical spin chain (\ref{eq:TFIM}), we introduce $\delta H=-\sum_{j=1}^{L}\mathcal{J}_{j}(t)n_{j}$ \cite{Bamba_2024_PRD}, where $n_{j}=(1-X_{j})/2$ is the local excitation-density operator. The source field is localized at $(t=0,j_{\text{origin}}=L/2)$, $\mathcal{J}_{j}(t)=\mathcal{A}\exp[-t^{2}/(2\sigma_{t}^{2})-\phi_{j}^{2}/(2\sigma_{\phi}^{2})-i\Omega t+iM\phi_{j}]$, with $\phi_{j}\equiv2\pi j/L-\pi$. Here $\sigma_{\phi}$ and $\sigma_{t}$ denote spatial and temporal widths, $M$ and $\Omega$ correspond to angular momentum and energy, and $\mathcal{A}=\sqrt{2L/(\sigma_{t}\sigma_{\phi})}/(4\pi)$. The antipodal transport signal at $j_{\mathrm{antipode}}=L$ is given by the linear response at time $t_{\mathrm{trans}}=\pi$, which equals the travel time of null geodesics between antipodal boundary points in pure AdS \cite{Sup_Mat}: $\delta\langle n_{j_{\mathrm{antipode}}}(t_{\mathrm{trans}})\rangle_{T}=-\sum_{j=1}^{L}\int_{-\infty}^{+\infty}dt\mathcal{J}_{j}(t)G_{R}(t_{\mathrm{trans}}-t,j_{\mathrm{antipode}}-j)$, where $G_{R}(t'-t,j'-j)=-i\theta(t'-t)\langle[n_{j'}(t'),n_{j}(t)]\rangle_{T}$ is the retarded propagator.

Because the BTZ contribution does not reach the boundary, the antipodal response is entirely determined by the AdS sector. The gravitational description therefore predicts the universal relation 
\begin{equation}
\frac{\left|\delta\langle n_{j_{\mathrm{antipode}}}(t_{\mathrm{trans}})\rangle_{T}\right|}{\left|\delta\langle n_{j_{\mathrm{antipode}}}(t_{\mathrm{trans}})\rangle_{T=0}\right|}=\frac{Z_{\mathrm{AdS}}(T)}{Z_{\mathrm{grav}}(T)},
\end{equation}
independent of the source profile. As shown in Fig.~\ref{Fig_1}(b), numerical results for different system sizes and angular momenta $M$ collapse onto the universal curve $Z_{\mathrm{AdS}}(T)/Z_{\mathrm{grav}}(T)=1/[1+e^{-1/(8G_{\mathrm{eff}}T)}e^{\pi^{2}\ell_{\mathrm{eff}}^{2}T/(2G_{\mathrm{eff}})}]$, with $\ell_{\mathrm{eff}}=1.28$ and $G_{\mathrm{eff}}=1.33$ \cite{Sup_Mat}, providing clear evidence for universal finite-temperature transport in the critical spin chain. 

Finally, the effective AdS radius and Newton constant deviate from the classical values $(\ell_{\mathrm{classical}}=1,G_{\mathrm{classical}}=3)$ obtained from the Brown--Henneaux analysis \cite{Brown_Henneaux_1986_CMP}. We attribute this to the small central charge $c=1/2$ of the Ising CFT, for which the bulk geometry is strongly quantum and higher-curvature corrections to the Einstein--Hilbert action become important \cite{Stelle_1977_PRD,Stelle_1978_GRG,Witten_1986_NPB,Emparan_2020_JHEP}. The collapse of the data onto the universal curve suggests that these quantum-gravity effects can be effectively captured through renormalized gravitational parameters $\ell_{\mathrm{eff}}$ and $G_{\mathrm{eff}}$, consistent with estimates from higher-curvature corrections \cite{Emparan_2020_JHEP}. 

\emph{Quasi-normal modes of the black hole in the critical spin chain.}---We now turn to the high-temperature regime, where the BTZ black hole dominates the effective gravitational description (\ref{eq:Partition_function_gravity_system}). In this regime, more direct signatures of black-hole physics are expected to emerge in the critical spin chain. We therefore focus on a prototypical dynamical feature of black holes: quasi-normal modes (QNMs) \cite{Konoplya_2011_RMP}. QNMs describe characteristic decaying oscillations of black-hole spacetime: after a perturbation is created outside the horizon, part of the disturbance is absorbed by the horizon and does not return. In the AdS/CFT correspondence, QNMs of the bulk black hole correspond to the relaxation of perturbations in the boundary CFT back to thermal equilibrium, with the relaxation rate determined by the imaginary part of the QNM frequency \cite{Birmingham_2002_PRL}. For the scalar perturbation $\delta H=-\sum_{j=1}^{L}\mathcal{J}_{j}(t)n_{j}$, introduced above, the lowest QNM is purely decaying with frequency $\omega_{{\rm QNM}}=-i2\pi T_{\mathrm{H}}\Delta$ \cite{Birmingham_2002_PRL,Janik_2025_PRD}. Here $T_{\mathrm{H}}=T$ is the Hawking temperature of the black hole, equal to the temperature of the dual CFT, and $\Delta=1$ is the scaling dimension of the operator $n_{j}$ in the Ising CFT \footnote{The lattice operator $n_{j}$ is not itself a primary operator, but in the scaling limit it flows to the CFT energy operator plus subleading corrections \cite{Zou_2020_PRL}, and therefore its long-distance correlations scale with dimension $\Delta=1$.}. This result predicts that, in the high-temperature regime, the linear response of the critical spin chain should exhibit exponential decay governed by this QNM. 

To test this prediction, we compute the spatially summed retarded response function $R(t)\equiv-i\theta(t)\sum_{j=1}^{L}\langle[n_{j}(t),n_{1}(0)]\rangle$. At sufficiently high temperature, the long-time behavior is expected to follow the QNM decay 
\begin{equation}
|R(t)|\propto e^{-i\omega_{{\rm QNM}}t}=e^{-(2\pi T\Delta)t}.
\end{equation}
Indeed, as shown in Fig.~\ref{Fig_2}, $|R(t)|$ exhibits a clear exponential decay over an extended time window. In addition to the leading exponential behavior, $|R(t)|$ displays a temperature-dependent offset $\exp(-\pi^{2}T/2)/2$, which can be attributed to loop corrections near the black-hole horizon \cite{Kraus_2017_JHEP,Janik_2025_PRD}. These results provide direct evidence that the relaxation dynamics of the critical spin chain at high temperature is governed by the quasi-normal modes of the dual BTZ black hole. 

\emph{Hawking--Page transition in the critical spin chain.}---The above results demonstrate that several characteristic features of black-hole physics manifest themselves in the critical spin chain. We therefore ask whether the Hawking--Page transition \cite{Hawking_Page_1983_CMP} in the dual gravitational description can also appear in the thermodynamics of the spin chain. From the gravitational partition function Eq.~(\ref{eq:Partition_function_gravity_system}), we see that in the low-temperature regime the AdS contribution $Z_{\mathrm{AdS}}(T)$ dominates, giving $S_{\mathrm{grav}}(T)\approx S_{\mathrm{AdS}}(T)=0$. In the high-temperature regime the BTZ contribution $Z_{\mathrm{BTZ}}(T)$ dominates, yielding $S_{\mathrm{grav}}(T)\approx S_{\mathrm{BTZ}}(T)\propto T$. This implies that the entropy growth rate should exhibit a characteristic feature at intermediate temperatures. Indeed, one finds that $dS_{\mathrm{grav}}(T)/dT$ develops a local minimum occurring precisely at the Hawking--Page temperature $T_{\mathrm{HP}}=1/(2\pi\ell)$, providing a concrete signal of the transition. 

To test this prediction, we compute the von Neumann entropy of the spin chain (\ref{eq:TFIM}) at the critical point $g=1$ \cite{Sup_Mat}, $S(T)=-\mathrm{tr}\left[\rho(T)\ln\rho(T)\right]$ with $\rho(T)\equiv e^{-H/T}/\mathrm{tr}(e^{-H/T})$. The resulting entropy $S(T)$ for $L=10^{3}$ is shown in Fig.~\ref{Fig_3}. At low temperature, $S(T)$ increases slowly and remains close to zero, consistent with the bulk prediction $S_{\mathrm{AdS}}(T)=0$. At high temperature, the entropy grows approximately linearly with $T$, consistent with the BTZ scaling $S_{\mathrm{BTZ}}(T)\propto T$ (purple circles and curve in Fig.~\ref{Fig_3}). In the intermediate regime, the derivative $dS(T)/dT$ exhibits a pronounced minimum at $T=0.16\pm0.01$. This agrees well with the Hawking--Page transition temperature $T_{\mathrm{HP}}=1/2\pi\approx0.16$, providing a clear thermodynamic signature of the Hawking--Page transition in the critical spin chain. 

\emph{Conclusion}.---We have demonstrated that the finite-temperature critical transverse-field Ising chain exhibits three quantitative and mutually consistent signatures of black-hole physics in its dual gravitational description within the AdS/CFT correspondence. First, antipodal excitation transport collapses onto a universal temperature-dependent curve determined by the relative weights of thermal AdS and the BTZ black hole saddle, directly reflecting horizon absorption. Second, in the high-temperature regime, the retarded response displays exponential decay governed by the lowest quasi-normal mode of the BTZ black hole. Third, the temperature derivative of the von Neumann entropy develops a pronounced local minimum at a temperature consistent with the Hawking--Page transition. Remarkably, these gravitational structures emerge in a system with minimal central charge. While holography is often discussed in the semiclassical large-$c$ limit, our results indicate that characteristic dynamical and thermodynamic features associated with black holes can persist beyond this regime. In this setting, the bulk description acts as an organizing principle that unifies transport, relaxation, and entropy across temperature scales. Given ongoing advances in high-precision quantum simulations of spin models, our results position critical quantum spin systems as experimentally accessible platforms for exploring dynamical and thermodynamic aspects of black-hole physics in controllable many-body experiments.
\begin{acknowledgments}
This work was supported by the NKRDPC (Grant No.~2022YFA1405304), the NSFC (Grant No.~12275089), Guangdong Basic and Applied Research Foundation (Grant No.~2023A1515012800). 
\end{acknowledgments}


\begin{thebibliography}{51}%
\makeatletter
\providecommand \@ifxundefined [1]{%
 \@ifx{#1\undefined}
}%
\providecommand \@ifnum [1]{%
 \ifnum #1\expandafter \@firstoftwo
 \else \expandafter \@secondoftwo
 \fi
}%
\providecommand \@ifx [1]{%
 \ifx #1\expandafter \@firstoftwo
 \else \expandafter \@secondoftwo
 \fi
}%
\providecommand \natexlab [1]{#1}%
\providecommand \enquote  [1]{``#1''}%
\providecommand \bibnamefont  [1]{#1}%
\providecommand \bibfnamefont [1]{#1}%
\providecommand \citenamefont [1]{#1}%
\providecommand \href@noop [0]{\@secondoftwo}%
\providecommand \href [0]{\begingroup \@sanitize@url \@href}%
\providecommand \@href[1]{\@@startlink{#1}\@@href}%
\providecommand \@@href[1]{\endgroup#1\@@endlink}%
\providecommand \@sanitize@url [0]{\catcode `\\12\catcode `\$12\catcode `\&12\catcode `\#12\catcode `\^12\catcode `\_12\catcode `\%12\relax}%
\providecommand \@@startlink[1]{}%
\providecommand \@@endlink[0]{}%
\providecommand \url  [0]{\begingroup\@sanitize@url \@url }%
\providecommand \@url [1]{\endgroup\@href {#1}{\urlprefix }}%
\providecommand \urlprefix  [0]{URL }%
\providecommand \Eprint [0]{\href }%
\providecommand \doibase [0]{https://doi.org/}%
\providecommand \selectlanguage [0]{\@gobble}%
\providecommand \bibinfo  [0]{\@secondoftwo}%
\providecommand \bibfield  [0]{\@secondoftwo}%
\providecommand \translation [1]{[#1]}%
\providecommand \BibitemOpen [0]{}%
\providecommand \bibitemStop [0]{}%
\providecommand \bibitemNoStop [0]{.\EOS\space}%
\providecommand \EOS [0]{\spacefactor3000\relax}%
\providecommand \BibitemShut  [1]{\csname bibitem#1\endcsname}%
\let\auto@bib@innerbib\@empty
\bibitem [{\citenamefont {Maldacena}(1999)}]{Maldacena1999IJTP}%
  \BibitemOpen
  \bibfield  {author} {\bibinfo {author} {\bibfnamefont {J.}~\bibnamefont {Maldacena}},\ }\bibfield  {title} {\bibinfo {title} {{The Large-N Limit of Superconformal Field Theories and Supergravity}},\ }\href {https://doi.org/10.1023/A:1026654312961} {\bibfield  {journal} {\bibinfo  {journal} {Int. J. Theor. Phys.}\ }\textbf {\bibinfo {volume} {38}},\ \bibinfo {pages} {1113} (\bibinfo {year} {1999})}\BibitemShut {NoStop}%
\bibitem [{\citenamefont {Gubser}\ \emph {et~al.}(1998)\citenamefont {Gubser}, \citenamefont {Klebanov},\ and\ \citenamefont {Polyakov}}]{Gubser_PLB_1998}%
  \BibitemOpen
  \bibfield  {author} {\bibinfo {author} {\bibfnamefont {S.}~\bibnamefont {Gubser}}, \bibinfo {author} {\bibfnamefont {I.}~\bibnamefont {Klebanov}},\ and\ \bibinfo {author} {\bibfnamefont {A.}~\bibnamefont {Polyakov}},\ }\bibfield  {title} {\bibinfo {title} {{Gauge Theory Correlators from Non-Critical String Theory}},\ }\href {https://doi.org/https://doi.org/10.1016/S0370-2693(98)00377-3} {\bibfield  {journal} {\bibinfo  {journal} {Phys. Lett. B}\ }\textbf {\bibinfo {volume} {428}},\ \bibinfo {pages} {105} (\bibinfo {year} {1998})}\BibitemShut {NoStop}%
\bibitem [{\citenamefont {Witten}(1998{\natexlab{a}})}]{Witten1998ATMP}%
  \BibitemOpen
  \bibfield  {author} {\bibinfo {author} {\bibfnamefont {E.}~\bibnamefont {Witten}},\ }\bibfield  {title} {\bibinfo {title} {{Anti-de Sitter Space and Holography}},\ }\href {https://doi.org/10.4310/ATMP.1998.v2.n2.a2} {\bibfield  {journal} {\bibinfo  {journal} {Adv. Theor. Math. Phys.}\ }\textbf {\bibinfo {volume} {2}},\ \bibinfo {pages} {253} (\bibinfo {year} {1998}{\natexlab{a}})}\BibitemShut {NoStop}%
\bibitem [{\citenamefont {Sachdev}\ and\ \citenamefont {Ye}(1993)}]{Sachdev_Ye_1993_PRL}%
  \BibitemOpen
  \bibfield  {author} {\bibinfo {author} {\bibfnamefont {S.}~\bibnamefont {Sachdev}}\ and\ \bibinfo {author} {\bibfnamefont {J.}~\bibnamefont {Ye}},\ }\bibfield  {title} {\bibinfo {title} {{Gapless Spin-Fluid Ground State in a Random Quantum Heisenberg Magnet}},\ }\href {https://doi.org/10.1103/PhysRevLett.70.3339} {\bibfield  {journal} {\bibinfo  {journal} {Phys. Rev. Lett.}\ }\textbf {\bibinfo {volume} {70}},\ \bibinfo {pages} {3339} (\bibinfo {year} {1993})}\BibitemShut {NoStop}%
\bibitem [{\citenamefont {Kitaev}(2015)}]{Kitaev_2015_talk}%
  \BibitemOpen
  \bibfield  {author} {\bibinfo {author} {\bibfnamefont {A.}~\bibnamefont {Kitaev}},\ }\href {http://online.kitp.ucsb.edu/online/entangled15/kitaev/} {\bibinfo {title} {{A Simple Model of Quantum Holography}}},\ \bibinfo {howpublished} {Talks at KITP, April 7, 2015 and May 27, 2015} (\bibinfo {year} {2015}),\ \bibinfo {note} {kavli Institute for Theoretical Physics (KITP)}\BibitemShut {NoStop}%
\bibitem [{\citenamefont {Maldacena}\ and\ \citenamefont {Stanford}(2016)}]{Maldacena_Stanford_2016_PRD}%
  \BibitemOpen
  \bibfield  {author} {\bibinfo {author} {\bibfnamefont {J.}~\bibnamefont {Maldacena}}\ and\ \bibinfo {author} {\bibfnamefont {D.}~\bibnamefont {Stanford}},\ }\bibfield  {title} {\bibinfo {title} {{Remarks on the Sachdev-Ye-Kitaev Model}},\ }\href {https://doi.org/10.1103/PhysRevD.94.106002} {\bibfield  {journal} {\bibinfo  {journal} {Phys. Rev. D}\ }\textbf {\bibinfo {volume} {94}},\ \bibinfo {pages} {106002} (\bibinfo {year} {2016})}\BibitemShut {NoStop}%
\bibitem [{\citenamefont {Jafferis}\ \emph {et~al.}(2022)\citenamefont {Jafferis}, \citenamefont {Zlokapa}, \citenamefont {Lykken}, \citenamefont {Kolchmeyer}, \citenamefont {Davis}, \citenamefont {Lauk}, \citenamefont {Neven},\ and\ \citenamefont {Spiropulu}}]{Jafferis_2022_nature}%
  \BibitemOpen
  \bibfield  {author} {\bibinfo {author} {\bibfnamefont {D.}~\bibnamefont {Jafferis}}, \bibinfo {author} {\bibfnamefont {A.}~\bibnamefont {Zlokapa}}, \bibinfo {author} {\bibfnamefont {J.~D.}\ \bibnamefont {Lykken}}, \bibinfo {author} {\bibfnamefont {D.~K.}\ \bibnamefont {Kolchmeyer}}, \bibinfo {author} {\bibfnamefont {S.~I.}\ \bibnamefont {Davis}}, \bibinfo {author} {\bibfnamefont {N.}~\bibnamefont {Lauk}}, \bibinfo {author} {\bibfnamefont {H.}~\bibnamefont {Neven}},\ and\ \bibinfo {author} {\bibfnamefont {M.}~\bibnamefont {Spiropulu}},\ }\bibfield  {title} {\bibinfo {title} {{Traversable Wormhole Dynamics on a Quantum Processor}},\ }\href {https://doi.org/10.1038/s41586-022-05424-3} {\bibfield  {journal} {\bibinfo  {journal} {Nature}\ }\textbf {\bibinfo {volume} {612}},\ \bibinfo {pages} {51} (\bibinfo {year} {2022})}\BibitemShut {NoStop}%
\bibitem [{\citenamefont {Xu}\ \emph {et~al.}(2020)\citenamefont {Xu}, \citenamefont {Susskind}, \citenamefont {Su},\ and\ \citenamefont {Swingle}}]{Susskind_Swingle_2020_arxiv}%
  \BibitemOpen
  \bibfield  {author} {\bibinfo {author} {\bibfnamefont {S.}~\bibnamefont {Xu}}, \bibinfo {author} {\bibfnamefont {L.}~\bibnamefont {Susskind}}, \bibinfo {author} {\bibfnamefont {Y.}~\bibnamefont {Su}},\ and\ \bibinfo {author} {\bibfnamefont {B.}~\bibnamefont {Swingle}},\ }\href {https://arxiv.org/abs/2008.02303} {\bibinfo {title} {{A Sparse Model of Quantum Holography}}} (\bibinfo {year} {2020}),\ \Eprint {https://arxiv.org/abs/2008.02303} {arXiv:2008.02303} \BibitemShut {NoStop}%
\bibitem [{\citenamefont {Castro}\ \emph {et~al.}(2012)\citenamefont {Castro}, \citenamefont {Gaberdiel}, \citenamefont {Hartman}, \citenamefont {Maloney},\ and\ \citenamefont {Volpato}}]{Castro_2012_PRD}%
  \BibitemOpen
  \bibfield  {author} {\bibinfo {author} {\bibfnamefont {A.}~\bibnamefont {Castro}}, \bibinfo {author} {\bibfnamefont {M.~R.}\ \bibnamefont {Gaberdiel}}, \bibinfo {author} {\bibfnamefont {T.}~\bibnamefont {Hartman}}, \bibinfo {author} {\bibfnamefont {A.}~\bibnamefont {Maloney}},\ and\ \bibinfo {author} {\bibfnamefont {R.}~\bibnamefont {Volpato}},\ }\bibfield  {title} {\bibinfo {title} {{Gravity Dual of the Ising Model}},\ }\href {https://doi.org/10.1103/PhysRevD.85.024032} {\bibfield  {journal} {\bibinfo  {journal} {Phys. Rev. D}\ }\textbf {\bibinfo {volume} {85}},\ \bibinfo {pages} {024032} (\bibinfo {year} {2012})}\BibitemShut {NoStop}%
\bibitem [{\citenamefont {Jian}\ \emph {et~al.}(2020)\citenamefont {Jian}, \citenamefont {Ludwig}, \citenamefont {Luo}, \citenamefont {Sun},\ and\ \citenamefont {Wang}}]{Jian_2020_JHEP}%
  \BibitemOpen
  \bibfield  {author} {\bibinfo {author} {\bibfnamefont {C.-M.}\ \bibnamefont {Jian}}, \bibinfo {author} {\bibfnamefont {A.~W.~W.}\ \bibnamefont {Ludwig}}, \bibinfo {author} {\bibfnamefont {Z.-X.}\ \bibnamefont {Luo}}, \bibinfo {author} {\bibfnamefont {H.-Y.}\ \bibnamefont {Sun}},\ and\ \bibinfo {author} {\bibfnamefont {Z.}~\bibnamefont {Wang}},\ }\bibfield  {title} {\bibinfo {title} {{Establishing Strongly-Coupled 3D {AdS} Quantum Gravity with {Ising} Dual Using All-Genus Partition Functions}},\ }\href {https://doi.org/10.1007/JHEP10(2020)129} {\bibfield  {journal} {\bibinfo  {journal} {J. High Energy Phys.}\ }\textbf {\bibinfo {volume} {2020}}\bibinfo  {number} { (10)},\ \bibinfo {pages} {129}}\BibitemShut {NoStop}%
\bibitem [{\citenamefont {Sachdev}(1999)}]{Sachdev_1999_book_QPT}%
  \BibitemOpen
\bibfield  {number} {  }\bibfield  {author} {\bibinfo {author} {\bibfnamefont {S.}~\bibnamefont {Sachdev}},\ }\bibfield  {title} {\bibinfo {title} {{Quantum Phase Transitions}},\ }\href {https://doi.org/10.1088/2058-7058/12/4/23} {\bibfield  {journal} {\bibinfo  {journal} {Phys. World}\ }\textbf {\bibinfo {volume} {12}},\ \bibinfo {pages} {33} (\bibinfo {year} {1999})}\BibitemShut {NoStop}%
\bibitem [{\citenamefont {Monroe}\ \emph {et~al.}(2021)\citenamefont {Monroe}, \citenamefont {Campbell}, \citenamefont {Duan}, \citenamefont {Gong}, \citenamefont {Gorshkov}, \citenamefont {Hess}, \citenamefont {Islam}, \citenamefont {Kim}, \citenamefont {Linke}, \citenamefont {Pagano}, \citenamefont {Richerme}, \citenamefont {Senko},\ and\ \citenamefont {Yao}}]{Monroe2021RMP}%
  \BibitemOpen
  \bibfield  {author} {\bibinfo {author} {\bibfnamefont {C.}~\bibnamefont {Monroe}}, \bibinfo {author} {\bibfnamefont {W.~C.}\ \bibnamefont {Campbell}}, \bibinfo {author} {\bibfnamefont {L.-M.}\ \bibnamefont {Duan}}, \bibinfo {author} {\bibfnamefont {Z.-X.}\ \bibnamefont {Gong}}, \bibinfo {author} {\bibfnamefont {A.~V.}\ \bibnamefont {Gorshkov}}, \bibinfo {author} {\bibfnamefont {P.~W.}\ \bibnamefont {Hess}}, \bibinfo {author} {\bibfnamefont {R.}~\bibnamefont {Islam}}, \bibinfo {author} {\bibfnamefont {K.}~\bibnamefont {Kim}}, \bibinfo {author} {\bibfnamefont {N.~M.}\ \bibnamefont {Linke}}, \bibinfo {author} {\bibfnamefont {G.}~\bibnamefont {Pagano}}, \bibinfo {author} {\bibfnamefont {P.}~\bibnamefont {Richerme}}, \bibinfo {author} {\bibfnamefont {C.}~\bibnamefont {Senko}},\ and\ \bibinfo {author} {\bibfnamefont {N.~Y.}\ \bibnamefont {Yao}},\ }\bibfield  {title} {\bibinfo {title} {{Programmable Quantum Simulations of Spin Systems with Trapped Ions}},\ }\href {https://doi.org/10.1103/RevModPhys.93.025001} {\bibfield  {journal} {\bibinfo  {journal} {Rev. Mod. Phys.}\ }\textbf {\bibinfo {volume} {93}},\ \bibinfo {pages} {025001} (\bibinfo {year} {2021})}\BibitemShut {NoStop}%
\bibitem [{\citenamefont {Haghshenas}\ \emph {et~al.}(2024)\citenamefont {Haghshenas}, \citenamefont {Chertkov}, \citenamefont {DeCross}, \citenamefont {Gatterman}, \citenamefont {Gerber}, \citenamefont {Gilmore}, \citenamefont {Gresh}, \citenamefont {Hewitt}, \citenamefont {Horst}, \citenamefont {Matheny}, \citenamefont {Mengle}, \citenamefont {Neyenhuis}, \citenamefont {Hayes},\ and\ \citenamefont {Foss-Feig}}]{haghshenas_PRL_2024}%
  \BibitemOpen
  \bibfield  {author} {\bibinfo {author} {\bibfnamefont {R.}~\bibnamefont {Haghshenas}}, \bibinfo {author} {\bibfnamefont {E.}~\bibnamefont {Chertkov}}, \bibinfo {author} {\bibfnamefont {M.}~\bibnamefont {DeCross}}, \bibinfo {author} {\bibfnamefont {T.~M.}\ \bibnamefont {Gatterman}}, \bibinfo {author} {\bibfnamefont {J.~A.}\ \bibnamefont {Gerber}}, \bibinfo {author} {\bibfnamefont {K.}~\bibnamefont {Gilmore}}, \bibinfo {author} {\bibfnamefont {D.}~\bibnamefont {Gresh}}, \bibinfo {author} {\bibfnamefont {N.}~\bibnamefont {Hewitt}}, \bibinfo {author} {\bibfnamefont {C.~V.}\ \bibnamefont {Horst}}, \bibinfo {author} {\bibfnamefont {M.}~\bibnamefont {Matheny}}, \bibinfo {author} {\bibfnamefont {T.}~\bibnamefont {Mengle}}, \bibinfo {author} {\bibfnamefont {B.}~\bibnamefont {Neyenhuis}}, \bibinfo {author} {\bibfnamefont {D.}~\bibnamefont {Hayes}},\ and\ \bibinfo {author} {\bibfnamefont {M.}~\bibnamefont {Foss-Feig}},\ }\bibfield  {title} {\bibinfo {title} {{Probing Critical States of Matter on a Digital Quantum Computer}},\ }\href {https://doi.org/10.1103/PhysRevLett.133.266502} {\bibfield  {journal} {\bibinfo  {journal} {Phys. Rev. Lett.}\ }\textbf {\bibinfo {volume} {133}},\ \bibinfo {pages} {266502} (\bibinfo {year} {2024})}\BibitemShut {NoStop}%
\bibitem [{\citenamefont {Bernien}\ \emph {et~al.}(2017)\citenamefont {Bernien}, \citenamefont {Schwartz}, \citenamefont {Keesling}, \citenamefont {Levine}, \citenamefont {Omran}, \citenamefont {Pichler}, \citenamefont {Choi}, \citenamefont {Zibrov}, \citenamefont {Endres}, \citenamefont {Greiner}, \citenamefont {Vuleti{\'c}},\ and\ \citenamefont {Lukin}}]{Bernien_Nature_2017}%
  \BibitemOpen
  \bibfield  {author} {\bibinfo {author} {\bibfnamefont {H.}~\bibnamefont {Bernien}}, \bibinfo {author} {\bibfnamefont {S.}~\bibnamefont {Schwartz}}, \bibinfo {author} {\bibfnamefont {A.}~\bibnamefont {Keesling}}, \bibinfo {author} {\bibfnamefont {H.}~\bibnamefont {Levine}}, \bibinfo {author} {\bibfnamefont {A.}~\bibnamefont {Omran}}, \bibinfo {author} {\bibfnamefont {H.}~\bibnamefont {Pichler}}, \bibinfo {author} {\bibfnamefont {S.}~\bibnamefont {Choi}}, \bibinfo {author} {\bibfnamefont {A.~S.}\ \bibnamefont {Zibrov}}, \bibinfo {author} {\bibfnamefont {M.}~\bibnamefont {Endres}}, \bibinfo {author} {\bibfnamefont {M.}~\bibnamefont {Greiner}}, \bibinfo {author} {\bibfnamefont {V.}~\bibnamefont {Vuleti{\'c}}},\ and\ \bibinfo {author} {\bibfnamefont {M.~D.}\ \bibnamefont {Lukin}},\ }\bibfield  {title} {\bibinfo {title} {{Probing Many-Body Dynamics on a 51-Atom Quantum Simulator}},\ }\href {https://doi.org/10.1038/nature24622} {\bibfield  {journal} {\bibinfo  {journal} {Nature}\ }\textbf {\bibinfo {volume} {551}},\ \bibinfo {pages} {579} (\bibinfo {year} {2017})}\BibitemShut {NoStop}%
\bibitem [{\citenamefont {Keesling}\ \emph {et~al.}(2019)\citenamefont {Keesling}, \citenamefont {Omran}, \citenamefont {Levine}, \citenamefont {Bernien}, \citenamefont {Pichler}, \citenamefont {Choi}, \citenamefont {Samajdar}, \citenamefont {Schwartz}, \citenamefont {Silvi}, \citenamefont {Sachdev}, \citenamefont {Zoller}, \citenamefont {Endres}, \citenamefont {Greiner}, \citenamefont {Vuleti{\'c}},\ and\ \citenamefont {Lukin}}]{Keesling_Nature_2019}%
  \BibitemOpen
  \bibfield  {author} {\bibinfo {author} {\bibfnamefont {A.}~\bibnamefont {Keesling}}, \bibinfo {author} {\bibfnamefont {A.}~\bibnamefont {Omran}}, \bibinfo {author} {\bibfnamefont {H.}~\bibnamefont {Levine}}, \bibinfo {author} {\bibfnamefont {H.}~\bibnamefont {Bernien}}, \bibinfo {author} {\bibfnamefont {H.}~\bibnamefont {Pichler}}, \bibinfo {author} {\bibfnamefont {S.}~\bibnamefont {Choi}}, \bibinfo {author} {\bibfnamefont {R.}~\bibnamefont {Samajdar}}, \bibinfo {author} {\bibfnamefont {S.}~\bibnamefont {Schwartz}}, \bibinfo {author} {\bibfnamefont {P.}~\bibnamefont {Silvi}}, \bibinfo {author} {\bibfnamefont {S.}~\bibnamefont {Sachdev}}, \bibinfo {author} {\bibfnamefont {P.}~\bibnamefont {Zoller}}, \bibinfo {author} {\bibfnamefont {M.}~\bibnamefont {Endres}}, \bibinfo {author} {\bibfnamefont {M.}~\bibnamefont {Greiner}}, \bibinfo {author} {\bibfnamefont {V.}~\bibnamefont {Vuleti{\'c}}},\ and\ \bibinfo {author} {\bibfnamefont {M.~D.}\ \bibnamefont {Lukin}},\ }\bibfield  {title} {\bibinfo {title} {Quantum {K}ibble--{Z}urek mechanism and critical dynamics on a programmable {R}ydberg simulator},\ }\href {https://doi.org/10.1038/s41586-019-1070-1} {\bibfield  {journal} {\bibinfo  {journal} {Nature}\ }\textbf {\bibinfo {volume} {568}},\ \bibinfo {pages} {207} (\bibinfo {year} {2019})}\BibitemShut {NoStop}%
\bibitem [{\citenamefont {Ebadi}\ \emph {et~al.}(2021)\citenamefont {Ebadi}, \citenamefont {Wang}, \citenamefont {Levine}, \citenamefont {Keesling}, \citenamefont {Semeghini}, \citenamefont {Omran}, \citenamefont {Bluvstein}, \citenamefont {Samajdar}, \citenamefont {Pichler}, \citenamefont {Ho}, \citenamefont {Choi}, \citenamefont {Sachdev}, \citenamefont {Greiner}, \citenamefont {Vuleti{\'c}},\ and\ \citenamefont {Lukin}}]{Ebadi2021nature}%
  \BibitemOpen
  \bibfield  {author} {\bibinfo {author} {\bibfnamefont {S.}~\bibnamefont {Ebadi}}, \bibinfo {author} {\bibfnamefont {T.~T.}\ \bibnamefont {Wang}}, \bibinfo {author} {\bibfnamefont {H.}~\bibnamefont {Levine}}, \bibinfo {author} {\bibfnamefont {A.}~\bibnamefont {Keesling}}, \bibinfo {author} {\bibfnamefont {G.}~\bibnamefont {Semeghini}}, \bibinfo {author} {\bibfnamefont {A.}~\bibnamefont {Omran}}, \bibinfo {author} {\bibfnamefont {D.}~\bibnamefont {Bluvstein}}, \bibinfo {author} {\bibfnamefont {R.}~\bibnamefont {Samajdar}}, \bibinfo {author} {\bibfnamefont {H.}~\bibnamefont {Pichler}}, \bibinfo {author} {\bibfnamefont {W.~W.}\ \bibnamefont {Ho}}, \bibinfo {author} {\bibfnamefont {S.}~\bibnamefont {Choi}}, \bibinfo {author} {\bibfnamefont {S.}~\bibnamefont {Sachdev}}, \bibinfo {author} {\bibfnamefont {M.}~\bibnamefont {Greiner}}, \bibinfo {author} {\bibfnamefont {V.}~\bibnamefont {Vuleti{\'c}}},\ and\ \bibinfo {author} {\bibfnamefont {M.~D.}\ \bibnamefont {Lukin}},\ }\bibfield  {title} {\bibinfo {title} {{Quantum Phases of Matter on a 256-Atom Programmable Quantum Simulator}},\ }\href {https://doi.org/10.1038/s41586-021-03582-4} {\bibfield  {journal} {\bibinfo  {journal} {Nature}\ }\textbf {\bibinfo {volume} {595}},\ \bibinfo {pages} {227} (\bibinfo {year} {2021})}\BibitemShut {NoStop}%
\bibitem [{\citenamefont {Scholl}\ \emph {et~al.}(2021)\citenamefont {Scholl}, \citenamefont {Schuler}, \citenamefont {Williams}, \citenamefont {Eberharter}, \citenamefont {Barredo}, \citenamefont {Schymik}, \citenamefont {Lienhard}, \citenamefont {Henry}, \citenamefont {Lang}, \citenamefont {Lahaye}, \citenamefont {L{\"a}uchli},\ and\ \citenamefont {Browaeys}}]{Scholl2021nature}%
  \BibitemOpen
  \bibfield  {author} {\bibinfo {author} {\bibfnamefont {P.}~\bibnamefont {Scholl}}, \bibinfo {author} {\bibfnamefont {M.}~\bibnamefont {Schuler}}, \bibinfo {author} {\bibfnamefont {H.~J.}\ \bibnamefont {Williams}}, \bibinfo {author} {\bibfnamefont {A.~A.}\ \bibnamefont {Eberharter}}, \bibinfo {author} {\bibfnamefont {D.}~\bibnamefont {Barredo}}, \bibinfo {author} {\bibfnamefont {K.-N.}\ \bibnamefont {Schymik}}, \bibinfo {author} {\bibfnamefont {V.}~\bibnamefont {Lienhard}}, \bibinfo {author} {\bibfnamefont {L.-P.}\ \bibnamefont {Henry}}, \bibinfo {author} {\bibfnamefont {T.~C.}\ \bibnamefont {Lang}}, \bibinfo {author} {\bibfnamefont {T.}~\bibnamefont {Lahaye}}, \bibinfo {author} {\bibfnamefont {A.~M.}\ \bibnamefont {L{\"a}uchli}},\ and\ \bibinfo {author} {\bibfnamefont {A.}~\bibnamefont {Browaeys}},\ }\bibfield  {title} {\bibinfo {title} {Quantum simulation of 2{D} antiferromagnets with hundreds of {R}ydberg atoms},\ }\href {https://doi.org/10.1038/s41586-021-03585-1} {\bibfield  {journal} {\bibinfo  {journal} {Nature}\ }\textbf {\bibinfo {volume} {595}},\ \bibinfo {pages} {233} (\bibinfo {year} {2021})}\BibitemShut {NoStop}%
\bibitem [{\citenamefont {Bluvstein}\ \emph {et~al.}(2022)\citenamefont {Bluvstein}, \citenamefont {Levine}, \citenamefont {Semeghini}, \citenamefont {Wang}, \citenamefont {Ebadi}, \citenamefont {Kalinowski}, \citenamefont {Keesling}, \citenamefont {Maskara}, \citenamefont {Pichler}, \citenamefont {Greiner}, \citenamefont {Vuleti{\'c}},\ and\ \citenamefont {Lukin}}]{Bluvstein_2022_nature}%
  \BibitemOpen
  \bibfield  {author} {\bibinfo {author} {\bibfnamefont {D.}~\bibnamefont {Bluvstein}}, \bibinfo {author} {\bibfnamefont {H.}~\bibnamefont {Levine}}, \bibinfo {author} {\bibfnamefont {G.}~\bibnamefont {Semeghini}}, \bibinfo {author} {\bibfnamefont {T.~T.}\ \bibnamefont {Wang}}, \bibinfo {author} {\bibfnamefont {S.}~\bibnamefont {Ebadi}}, \bibinfo {author} {\bibfnamefont {M.}~\bibnamefont {Kalinowski}}, \bibinfo {author} {\bibfnamefont {A.}~\bibnamefont {Keesling}}, \bibinfo {author} {\bibfnamefont {N.}~\bibnamefont {Maskara}}, \bibinfo {author} {\bibfnamefont {H.}~\bibnamefont {Pichler}}, \bibinfo {author} {\bibfnamefont {M.}~\bibnamefont {Greiner}}, \bibinfo {author} {\bibfnamefont {V.}~\bibnamefont {Vuleti{\'c}}},\ and\ \bibinfo {author} {\bibfnamefont {M.~D.}\ \bibnamefont {Lukin}},\ }\bibfield  {title} {\bibinfo {title} {{A Quantum Processor Based on Coherent Transport of Entangled Atom Arrays}},\ }\href {https://doi.org/10.1038/s41586-022-04592-6} {\bibfield  {journal} {\bibinfo  {journal} {Nature}\ }\textbf {\bibinfo {volume} {604}},\ \bibinfo {pages} {451} (\bibinfo {year} {2022})}\BibitemShut {NoStop}%
\bibitem [{\citenamefont {Fang}\ \emph {et~al.}()\citenamefont {Fang}, \citenamefont {Wang}, \citenamefont {Liu}, \citenamefont {Wang}, \citenamefont {Cimmino}, \citenamefont {Wei}, \citenamefont {Bintz}, \citenamefont {Parr}, \citenamefont {Kemp}, \citenamefont {Ni},\ and\ \citenamefont {Yao}}]{fang2024arxiv}%
  \BibitemOpen
  \bibfield  {author} {\bibinfo {author} {\bibfnamefont {F.}~\bibnamefont {Fang}}, \bibinfo {author} {\bibfnamefont {K.}~\bibnamefont {Wang}}, \bibinfo {author} {\bibfnamefont {V.~S.}\ \bibnamefont {Liu}}, \bibinfo {author} {\bibfnamefont {Y.}~\bibnamefont {Wang}}, \bibinfo {author} {\bibfnamefont {R.}~\bibnamefont {Cimmino}}, \bibinfo {author} {\bibfnamefont {J.}~\bibnamefont {Wei}}, \bibinfo {author} {\bibfnamefont {M.}~\bibnamefont {Bintz}}, \bibinfo {author} {\bibfnamefont {A.}~\bibnamefont {Parr}}, \bibinfo {author} {\bibfnamefont {J.}~\bibnamefont {Kemp}}, \bibinfo {author} {\bibfnamefont {K.-K.}\ \bibnamefont {Ni}},\ and\ \bibinfo {author} {\bibfnamefont {N.~Y.}\ \bibnamefont {Yao}},\ }\href@noop {} {\bibinfo {title} {{Probing Critical Phenomena in Open Quantum Systems Using Atom Arrays}}},\ \bibinfo {howpublished} {\href{https://arxiv.org/abs/2402.15376}{arXiv:2402.15376}}\BibitemShut {NoStop}%
\bibitem [{\citenamefont {Kim}\ \emph {et~al.}()\citenamefont {Kim}, \citenamefont {Lukin}, \citenamefont {Rispoli}, \citenamefont {Tai}, \citenamefont {Kaufman}, \citenamefont {Segura}, \citenamefont {Li}, \citenamefont {Kwan}, \citenamefont {L\'eonard}, \citenamefont {Bakkali-Hassani},\ and\ \citenamefont {Greiner}}]{kim2024arXiv}%
  \BibitemOpen
  \bibfield  {author} {\bibinfo {author} {\bibfnamefont {S.}~\bibnamefont {Kim}}, \bibinfo {author} {\bibfnamefont {A.}~\bibnamefont {Lukin}}, \bibinfo {author} {\bibfnamefont {M.}~\bibnamefont {Rispoli}}, \bibinfo {author} {\bibfnamefont {M.~E.}\ \bibnamefont {Tai}}, \bibinfo {author} {\bibfnamefont {A.~M.}\ \bibnamefont {Kaufman}}, \bibinfo {author} {\bibfnamefont {P.}~\bibnamefont {Segura}}, \bibinfo {author} {\bibfnamefont {Y.}~\bibnamefont {Li}}, \bibinfo {author} {\bibfnamefont {J.}~\bibnamefont {Kwan}}, \bibinfo {author} {\bibfnamefont {J.}~\bibnamefont {L\'eonard}}, \bibinfo {author} {\bibfnamefont {B.}~\bibnamefont {Bakkali-Hassani}},\ and\ \bibinfo {author} {\bibfnamefont {M.}~\bibnamefont {Greiner}},\ }\href@noop {} {\bibinfo {title} {Adiabatic state preparation in a quantum {I}sing spin chain}},\ \bibinfo {howpublished} {\href{https://arxiv.org/abs/2404.07481}{arXiv:2404.07481}}\BibitemShut {NoStop}%
\bibitem [{\citenamefont {Hawking}\ and\ \citenamefont {Page}(1983)}]{Hawking_Page_1983_CMP}%
  \BibitemOpen
  \bibfield  {author} {\bibinfo {author} {\bibfnamefont {S.~W.}\ \bibnamefont {Hawking}}\ and\ \bibinfo {author} {\bibfnamefont {D.~N.}\ \bibnamefont {Page}},\ }\bibfield  {title} {\bibinfo {title} {{Thermodynamics of Black Holes in Anti-de Sitter Space}},\ }\href {https://doi.org/10.1007/BF01208266} {\bibfield  {journal} {\bibinfo  {journal} {Commun. Math. Phys.}\ }\textbf {\bibinfo {volume} {87}},\ \bibinfo {pages} {577} (\bibinfo {year} {1983})}\BibitemShut {NoStop}%
\bibitem [{\citenamefont {Witten}(1998{\natexlab{b}})}]{Witten_1998_ads_confinement}%
  \BibitemOpen
  \bibfield  {author} {\bibinfo {author} {\bibfnamefont {E.}~\bibnamefont {Witten}},\ }\href {https://arxiv.org/abs/hep-th/9803131} {\bibinfo {title} {{Anti-de Sitter Space, Thermal Phase Transition, and Confinement in Gauge Theories}}} (\bibinfo {year} {1998}{\natexlab{b}}),\ \Eprint {https://arxiv.org/abs/hep-th/9803131} {arXiv:hep-th/9803131} \BibitemShut {NoStop}%
\bibitem [{\citenamefont {Birmingham}\ \emph {et~al.}(2002)\citenamefont {Birmingham}, \citenamefont {Sachs},\ and\ \citenamefont {Solodukhin}}]{Birmingham_2002_PRL}%
  \BibitemOpen
  \bibfield  {author} {\bibinfo {author} {\bibfnamefont {D.}~\bibnamefont {Birmingham}}, \bibinfo {author} {\bibfnamefont {I.}~\bibnamefont {Sachs}},\ and\ \bibinfo {author} {\bibfnamefont {S.~N.}\ \bibnamefont {Solodukhin}},\ }\bibfield  {title} {\bibinfo {title} {{Conformal Field Theory Interpretation of Black Hole Quasinormal Modes}},\ }\href {https://doi.org/10.1103/PhysRevLett.88.151301} {\bibfield  {journal} {\bibinfo  {journal} {Phys. Rev. Lett.}\ }\textbf {\bibinfo {volume} {88}},\ \bibinfo {pages} {151301} (\bibinfo {year} {2002})}\BibitemShut {NoStop}%
\bibitem [{\citenamefont {Konoplya}\ and\ \citenamefont {Zhidenko}(2011)}]{Konoplya_2011_RMP}%
  \BibitemOpen
  \bibfield  {author} {\bibinfo {author} {\bibfnamefont {R.~A.}\ \bibnamefont {Konoplya}}\ and\ \bibinfo {author} {\bibfnamefont {A.}~\bibnamefont {Zhidenko}},\ }\bibfield  {title} {\bibinfo {title} {{Quasinormal Modes of Black Holes: From Astrophysics to String Theory}},\ }\href {https://doi.org/10.1103/RevModPhys.83.793} {\bibfield  {journal} {\bibinfo  {journal} {Rev. Mod. Phys.}\ }\textbf {\bibinfo {volume} {83}},\ \bibinfo {pages} {793} (\bibinfo {year} {2011})}\BibitemShut {NoStop}%
\bibitem [{\citenamefont {Francesco}\ \emph {et~al.}(2012)\citenamefont {Francesco}, \citenamefont {Mathieu},\ and\ \citenamefont {S{\'e}n{\'e}chal}}]{Francesco_CFT_Book_2012}%
  \BibitemOpen
  \bibfield  {author} {\bibinfo {author} {\bibfnamefont {P.}~\bibnamefont {Francesco}}, \bibinfo {author} {\bibfnamefont {P.}~\bibnamefont {Mathieu}},\ and\ \bibinfo {author} {\bibfnamefont {D.}~\bibnamefont {S{\'e}n{\'e}chal}},\ }\href@noop {} {\emph {\bibinfo {title} {{Conformal Field Theory}}}}\ (\bibinfo  {publisher} {Springer New York, NY},\ \bibinfo {year} {2012})\BibitemShut {NoStop}%
\bibitem [{\citenamefont {Bamba}\ \emph {et~al.}(2024)\citenamefont {Bamba}, \citenamefont {Hashimoto}, \citenamefont {Murata}, \citenamefont {Takeda},\ and\ \citenamefont {Yamamoto}}]{Bamba_2024_PRD}%
  \BibitemOpen
  \bibfield  {author} {\bibinfo {author} {\bibfnamefont {M.}~\bibnamefont {Bamba}}, \bibinfo {author} {\bibfnamefont {K.}~\bibnamefont {Hashimoto}}, \bibinfo {author} {\bibfnamefont {K.}~\bibnamefont {Murata}}, \bibinfo {author} {\bibfnamefont {D.}~\bibnamefont {Takeda}},\ and\ \bibinfo {author} {\bibfnamefont {D.}~\bibnamefont {Yamamoto}},\ }\bibfield  {title} {\bibinfo {title} {{Spacetime-Localized Response in Quantum Critical Spin Systems: Insights from Holography}},\ }\href {https://doi.org/10.1103/PhysRevD.109.126003} {\bibfield  {journal} {\bibinfo  {journal} {Phys. Rev. D}\ }\textbf {\bibinfo {volume} {109}},\ \bibinfo {pages} {126003} (\bibinfo {year} {2024})}\BibitemShut {NoStop}%
\bibitem [{\citenamefont {Sahay}\ \emph {et~al.}(2025)\citenamefont {Sahay}, \citenamefont {Lukin},\ and\ \citenamefont {Cotler}}]{Sahay_2025_PRX}%
  \BibitemOpen
  \bibfield  {author} {\bibinfo {author} {\bibfnamefont {R.}~\bibnamefont {Sahay}}, \bibinfo {author} {\bibfnamefont {M.~D.}\ \bibnamefont {Lukin}},\ and\ \bibinfo {author} {\bibfnamefont {J.}~\bibnamefont {Cotler}},\ }\bibfield  {title} {\bibinfo {title} {{Emergent Holographic Forces from Tensor Networks and Criticality}},\ }\href {https://doi.org/10.1103/PhysRevX.15.021078} {\bibfield  {journal} {\bibinfo  {journal} {Phys. Rev. X}\ }\textbf {\bibinfo {volume} {15}},\ \bibinfo {pages} {021078} (\bibinfo {year} {2025})}\BibitemShut {NoStop}%
\bibitem [{\citenamefont {Imagawa}\ \emph {et~al.}()\citenamefont {Imagawa}, \citenamefont {Murata},\ and\ \citenamefont {Yamamoto}}]{imagawa2025arXiv}%
  \BibitemOpen
  \bibfield  {author} {\bibinfo {author} {\bibfnamefont {D.}~\bibnamefont {Imagawa}}, \bibinfo {author} {\bibfnamefont {K.}~\bibnamefont {Murata}},\ and\ \bibinfo {author} {\bibfnamefont {D.}~\bibnamefont {Yamamoto}},\ }\href@noop {} {\bibinfo {title} {{Operator Dependence and Robustness of Spacetime-Localized Response in a Quantum Critical Spin Chain}}},\ \bibinfo {howpublished} {\href{https://arxiv.org/abs/2510.04047}{arXiv:2510.04047}}\BibitemShut {NoStop}%
\bibitem [{\citenamefont {Wang}\ and\ \citenamefont {He}(2025)}]{Wang_2025_arXiv}%
  \BibitemOpen
  \bibfield  {author} {\bibinfo {author} {\bibfnamefont {Z.}~\bibnamefont {Wang}}\ and\ \bibinfo {author} {\bibfnamefont {L.}~\bibnamefont {He}},\ }\href {https://arxiv.org/abs/2501.03084} {\bibinfo {title} {{Stable Excitations and Holographic Transportation in Tensor Networks of Critical Spin Chains}}} (\bibinfo {year} {2025}),\ \Eprint {https://arxiv.org/abs/2501.03084} {arXiv:2501.03084} \BibitemShut {NoStop}%
\bibitem [{\citenamefont {Kurita}\ and\ \citenamefont {Sakagami}(2005)}]{Kurita_2005_PTP}%
  \BibitemOpen
  \bibfield  {author} {\bibinfo {author} {\bibfnamefont {Y.}~\bibnamefont {Kurita}}\ and\ \bibinfo {author} {\bibfnamefont {M.-a.}\ \bibnamefont {Sakagami}},\ }\bibfield  {title} {\bibinfo {title} {{CFT Description of the Three-Dimensional Hawking-Page Transition}},\ }\href {https://doi.org/10.1143/PTP.113.1193} {\bibfield  {journal} {\bibinfo  {journal} {Prog. Theor. Phys.}\ }\textbf {\bibinfo {volume} {113}},\ \bibinfo {pages} {1193} (\bibinfo {year} {2005})}\BibitemShut {NoStop}%
\bibitem [{\citenamefont {Witten}(2024)}]{Witten_2024_arxiv_introductionblackholethermodynamics}%
  \BibitemOpen
  \bibfield  {author} {\bibinfo {author} {\bibfnamefont {E.}~\bibnamefont {Witten}},\ }\href {https://arxiv.org/abs/2412.16795} {\bibinfo {title} {{Introduction to Black Hole Thermodynamics}}} (\bibinfo {year} {2024}),\ \Eprint {https://arxiv.org/abs/2412.16795} {arXiv:2412.16795} \BibitemShut {NoStop}%
\bibitem [{\citenamefont {Janik}(2025)}]{Janik_2025_PRD}%
  \BibitemOpen
  \bibfield  {author} {\bibinfo {author} {\bibfnamefont {R.~A.}\ \bibnamefont {Janik}},\ }\bibfield  {title} {\bibinfo {title} {{Ising Model as a Window on Quantum Gravity with Matter}},\ }\href {https://doi.org/10.1103/PhysRevD.111.106016} {\bibfield  {journal} {\bibinfo  {journal} {Phys. Rev. D}\ }\textbf {\bibinfo {volume} {111}},\ \bibinfo {pages} {106016} (\bibinfo {year} {2025})}\BibitemShut {NoStop}%
\bibitem [{\citenamefont {Nguyen}\ \emph {et~al.}(2018)\citenamefont {Nguyen}, \citenamefont {Devakul}, \citenamefont {Halbasch}, \citenamefont {Zaletel},\ and\ \citenamefont {Swingle}}]{Nguyen_Brian_2018_JHEP}%
  \BibitemOpen
  \bibfield  {author} {\bibinfo {author} {\bibfnamefont {P.}~\bibnamefont {Nguyen}}, \bibinfo {author} {\bibfnamefont {T.}~\bibnamefont {Devakul}}, \bibinfo {author} {\bibfnamefont {M.~G.}\ \bibnamefont {Halbasch}}, \bibinfo {author} {\bibfnamefont {M.~P.}\ \bibnamefont {Zaletel}},\ and\ \bibinfo {author} {\bibfnamefont {B.}~\bibnamefont {Swingle}},\ }\bibfield  {title} {\bibinfo {title} {{Entanglement of Purification: From Spin Chains to Holography}},\ }\href {https://doi.org/10.1007/JHEP01(2018)098} {\bibfield  {journal} {\bibinfo  {journal} {J. High Energy Phys.}\ }\textbf {\bibinfo {volume} {2018}}\bibinfo  {number} { (1)},\ \bibinfo {pages} {98}}\BibitemShut {NoStop}%
\bibitem [{\citenamefont {P\'erez-Garc\'{\i}a}\ \emph {et~al.}(2024)\citenamefont {P\'erez-Garc\'{\i}a}, \citenamefont {Santilli},\ and\ \citenamefont {Tierz}}]{Miguel_2024_PRR}%
  \BibitemOpen
\bibfield  {number} {  }\bibfield  {author} {\bibinfo {author} {\bibfnamefont {D.}~\bibnamefont {P\'erez-Garc\'{\i}a}}, \bibinfo {author} {\bibfnamefont {L.}~\bibnamefont {Santilli}},\ and\ \bibinfo {author} {\bibfnamefont {M.}~\bibnamefont {Tierz}},\ }\bibfield  {title} {\bibinfo {title} {{Hawking-Page Transition on a Spin Chain}},\ }\href {https://doi.org/10.1103/PhysRevResearch.6.033007} {\bibfield  {journal} {\bibinfo  {journal} {Phys. Rev. Res.}\ }\textbf {\bibinfo {volume} {6}},\ \bibinfo {pages} {033007} (\bibinfo {year} {2024})}\BibitemShut {NoStop}%
\bibitem [{\citenamefont {Ba\~nados}\ \emph {et~al.}(1992)\citenamefont {Ba\~nados}, \citenamefont {Teitelboim},\ and\ \citenamefont {Zanelli}}]{Zanelli_1992_PRL}%
  \BibitemOpen
  \bibfield  {author} {\bibinfo {author} {\bibfnamefont {M.}~\bibnamefont {Ba\~nados}}, \bibinfo {author} {\bibfnamefont {C.}~\bibnamefont {Teitelboim}},\ and\ \bibinfo {author} {\bibfnamefont {J.}~\bibnamefont {Zanelli}},\ }\bibfield  {title} {\bibinfo {title} {{Black Hole in Three-Dimensional Spacetime}},\ }\href {https://doi.org/10.1103/PhysRevLett.69.1849} {\bibfield  {journal} {\bibinfo  {journal} {Phys. Rev. Lett.}\ }\textbf {\bibinfo {volume} {69}},\ \bibinfo {pages} {1849} (\bibinfo {year} {1992})}\BibitemShut {NoStop}%
\bibitem [{\citenamefont {Ba\~nados}\ \emph {et~al.}(1993)\citenamefont {Ba\~nados}, \citenamefont {Henneaux}, \citenamefont {Teitelboim},\ and\ \citenamefont {Zanelli}}]{Zanelli_1993_PRD}%
  \BibitemOpen
  \bibfield  {author} {\bibinfo {author} {\bibfnamefont {M.}~\bibnamefont {Ba\~nados}}, \bibinfo {author} {\bibfnamefont {M.}~\bibnamefont {Henneaux}}, \bibinfo {author} {\bibfnamefont {C.}~\bibnamefont {Teitelboim}},\ and\ \bibinfo {author} {\bibfnamefont {J.}~\bibnamefont {Zanelli}},\ }\bibfield  {title} {\bibinfo {title} {{Geometry of the 2+1 Black Hole}},\ }\href {https://doi.org/10.1103/PhysRevD.48.1506} {\bibfield  {journal} {\bibinfo  {journal} {Phys. Rev. D}\ }\textbf {\bibinfo {volume} {48}},\ \bibinfo {pages} {1506} (\bibinfo {year} {1993})}\BibitemShut {NoStop}%
\bibitem [{Sup()}]{Sup_Mat}%
  \BibitemOpen
  \href@noop {} {\bibinfo  {journal} {See Supplemental Material for a discussion on relevant technical details}\ }\BibitemShut {NoStop}%
\bibitem [{\citenamefont {{Frolov}}\ and\ \citenamefont {{Novikov}}(1998)}]{Frolov_Black_hole_Physics_Book_1998}%
  \BibitemOpen
\bibfield  {journal} {  }\bibfield  {author} {\bibinfo {author} {\bibfnamefont {V.~P.}\ \bibnamefont {{Frolov}}}\ and\ \bibinfo {author} {\bibfnamefont {I.~D.}\ \bibnamefont {{Novikov}}},\ }\href@noop {} {\emph {\bibinfo {title} {{Black Hole Physics: Basic Concepts and New Developments}}}}\ (\bibinfo  {publisher} {Springer Dordrecht},\ \bibinfo {year} {1998})\BibitemShut {NoStop}%
\bibitem [{\citenamefont {Kinoshita}\ \emph {et~al.}(2023)\citenamefont {Kinoshita}, \citenamefont {Murata},\ and\ \citenamefont {Takeda}}]{Kinoshita_2023_JHEP}%
  \BibitemOpen
  \bibfield  {author} {\bibinfo {author} {\bibfnamefont {S.}~\bibnamefont {Kinoshita}}, \bibinfo {author} {\bibfnamefont {K.}~\bibnamefont {Murata}},\ and\ \bibinfo {author} {\bibfnamefont {D.}~\bibnamefont {Takeda}},\ }\bibfield  {title} {\bibinfo {title} {{Shooting Null Geodesics into Holographic Spacetimes}},\ }\href {https://doi.org/10.1007/JHEP10(2023)074} {\bibfield  {journal} {\bibinfo  {journal} {J. High Energy Phys.}\ }\textbf {\bibinfo {volume} {2023}}\bibinfo  {number} { (10)},\ \bibinfo {pages} {74}}\BibitemShut {NoStop}%
\bibitem [{\citenamefont {Brown}\ and\ \citenamefont {Henneaux}(1986)}]{Brown_Henneaux_1986_CMP}%
  \BibitemOpen
\bibfield  {number} {  }\bibfield  {author} {\bibinfo {author} {\bibfnamefont {J.~D.}\ \bibnamefont {Brown}}\ and\ \bibinfo {author} {\bibfnamefont {M.}~\bibnamefont {Henneaux}},\ }\bibfield  {title} {\bibinfo {title} {{Central Charges in the Canonical Realization of Asymptotic Symmetries: An Example from Three Dimensional Gravity}},\ }\href {https://doi.org/10.1007/BF01211590} {\bibfield  {journal} {\bibinfo  {journal} {Commun. Math. Phys.}\ }\textbf {\bibinfo {volume} {104}},\ \bibinfo {pages} {207} (\bibinfo {year} {1986})}\BibitemShut {NoStop}%
\bibitem [{\citenamefont {Stelle}(1977)}]{Stelle_1977_PRD}%
  \BibitemOpen
  \bibfield  {author} {\bibinfo {author} {\bibfnamefont {K.~S.}\ \bibnamefont {Stelle}},\ }\bibfield  {title} {\bibinfo {title} {{Renormalization of Higher-Derivative Quantum Gravity}},\ }\href {https://doi.org/10.1103/PhysRevD.16.953} {\bibfield  {journal} {\bibinfo  {journal} {Phys. Rev. D}\ }\textbf {\bibinfo {volume} {16}},\ \bibinfo {pages} {953} (\bibinfo {year} {1977})}\BibitemShut {NoStop}%
\bibitem [{\citenamefont {Stelle}(1978)}]{Stelle_1978_GRG}%
  \BibitemOpen
  \bibfield  {author} {\bibinfo {author} {\bibfnamefont {K.~S.}\ \bibnamefont {Stelle}},\ }\bibfield  {title} {\bibinfo {title} {{Classical Gravity with Higher Derivatives}},\ }\href {https://doi.org/10.1007/BF00760427} {\bibfield  {journal} {\bibinfo  {journal} {Gen. Relativ. Gravit.}\ }\textbf {\bibinfo {volume} {9}},\ \bibinfo {pages} {353} (\bibinfo {year} {1978})}\BibitemShut {NoStop}%
\bibitem [{\citenamefont {Gross}\ and\ \citenamefont {Witten}(1986)}]{Witten_1986_NPB}%
  \BibitemOpen
  \bibfield  {author} {\bibinfo {author} {\bibfnamefont {D.~J.}\ \bibnamefont {Gross}}\ and\ \bibinfo {author} {\bibfnamefont {E.}~\bibnamefont {Witten}},\ }\bibfield  {title} {\bibinfo {title} {{Superstring Modifications of Einstein's Equations}},\ }\href {https://doi.org/https://doi.org/10.1016/0550-3213(86)90429-3} {\bibfield  {journal} {\bibinfo  {journal} {Nucl. Phys. B}\ }\textbf {\bibinfo {volume} {277}},\ \bibinfo {pages} {1} (\bibinfo {year} {1986})}\BibitemShut {NoStop}%
\bibitem [{\citenamefont {Emparan}\ \emph {et~al.}(2020)\citenamefont {Emparan}, \citenamefont {Frassino},\ and\ \citenamefont {Way}}]{Emparan_2020_JHEP}%
  \BibitemOpen
  \bibfield  {author} {\bibinfo {author} {\bibfnamefont {R.}~\bibnamefont {Emparan}}, \bibinfo {author} {\bibfnamefont {A.~M.}\ \bibnamefont {Frassino}},\ and\ \bibinfo {author} {\bibfnamefont {B.}~\bibnamefont {Way}},\ }\bibfield  {title} {\bibinfo {title} {{Quantum BTZ Black Hole}},\ }\href {https://doi.org/10.1007/JHEP11(2020)137} {\bibfield  {journal} {\bibinfo  {journal} {J. High Energy Phys.}\ }\textbf {\bibinfo {volume} {2020}}\bibinfo  {number} { (11)},\ \bibinfo {pages} {137}}\BibitemShut {NoStop}%
\bibitem [{Note1()}]{Note1}%
  \BibitemOpen
\bibfield  {number} {  }\bibinfo {note} {The lattice operator $n_{j}$ is not itself a primary operator, but in the scaling limit it flows to the CFT energy operator plus subleading corrections \cite {Zou_2020_PRL}, and therefore its long-distance correlations scale with dimension $\Delta =1$.}\BibitemShut {Stop}%
\bibitem [{\citenamefont {Kraus}\ and\ \citenamefont {Maloney}(2017)}]{Kraus_2017_JHEP}%
  \BibitemOpen
  \bibfield  {author} {\bibinfo {author} {\bibfnamefont {P.}~\bibnamefont {Kraus}}\ and\ \bibinfo {author} {\bibfnamefont {A.}~\bibnamefont {Maloney}},\ }\bibfield  {title} {\bibinfo {title} {{A Cardy Formula for Three-Point Coefficients or How the Black Hole Got Its Spots}},\ }\href {https://doi.org/10.1007/JHEP05(2017)160} {\bibfield  {journal} {\bibinfo  {journal} {J. High Energy Phys.}\ }\textbf {\bibinfo {volume} {2017}}\bibinfo  {number} { (5)},\ \bibinfo {pages} {160}}\BibitemShut {NoStop}%
\bibitem [{\citenamefont {Zou}\ \emph {et~al.}(2020)\citenamefont {Zou}, \citenamefont {Milsted},\ and\ \citenamefont {Vidal}}]{Zou_2020_PRL}%
  \BibitemOpen
\bibfield  {number} {  }\bibfield  {author} {\bibinfo {author} {\bibfnamefont {Y.}~\bibnamefont {Zou}}, \bibinfo {author} {\bibfnamefont {A.}~\bibnamefont {Milsted}},\ and\ \bibinfo {author} {\bibfnamefont {G.}~\bibnamefont {Vidal}},\ }\bibfield  {title} {\bibinfo {title} {{Conformal Fields and Operator Product Expansion in Critical Quantum Spin Chains}},\ }\href {https://doi.org/10.1103/PhysRevLett.124.040604} {\bibfield  {journal} {\bibinfo  {journal} {Phys. Rev. Lett.}\ }\textbf {\bibinfo {volume} {124}},\ \bibinfo {pages} {040604} (\bibinfo {year} {2020})}\BibitemShut {NoStop}%
\bibitem [{\citenamefont {Jordan}\ and\ \citenamefont {Wigner}(1928)}]{Jordan_Wigner_1928}%
  \BibitemOpen
  \bibfield  {author} {\bibinfo {author} {\bibfnamefont {P.}~\bibnamefont {Jordan}}\ and\ \bibinfo {author} {\bibfnamefont {E.}~\bibnamefont {Wigner}},\ }\bibfield  {title} {\bibinfo {title} {{\"U}ber das paulische {\"a}quivalenzverbot},\ }\href {https://doi.org/10.1007/BF01331938} {\bibfield  {journal} {\bibinfo  {journal} {Z. Phys.}\ }\textbf {\bibinfo {volume} {47}},\ \bibinfo {pages} {631} (\bibinfo {year} {1928})}\BibitemShut {NoStop}%
\bibitem [{\citenamefont {Weinberg}\ and\ \citenamefont {Bukov}(2019)}]{Phillip_2019_scipost}%
  \BibitemOpen
  \bibfield  {author} {\bibinfo {author} {\bibfnamefont {P.}~\bibnamefont {Weinberg}}\ and\ \bibinfo {author} {\bibfnamefont {M.}~\bibnamefont {Bukov}},\ }\bibfield  {title} {\bibinfo {title} {{QuSpin: A Python Package for Dynamics and Exact Diagonalisation of Quantum Many Body Systems. Part II: Bosons, Fermions and Higher Spins}},\ }\href {https://doi.org/10.21468/SciPostPhys.7.2.020} {\bibfield  {journal} {\bibinfo  {journal} {SciPost Phys.}\ }\textbf {\bibinfo {volume} {7}},\ \bibinfo {pages} {020} (\bibinfo {year} {2019})}\BibitemShut {NoStop}%
\bibitem [{\citenamefont {Wu}(2020)}]{Wu_2020_PRE}%
  \BibitemOpen
  \bibfield  {author} {\bibinfo {author} {\bibfnamefont {N.}~\bibnamefont {Wu}},\ }\bibfield  {title} {\bibinfo {title} {{Longitudinal Magnetization Dynamics in the Quantum Ising Ring: A Pfaffian Method Based on Correspondence Between Momentum Space and Real Space}},\ }\href {https://doi.org/10.1103/PhysRevE.101.042108} {\bibfield  {journal} {\bibinfo  {journal} {Phys. Rev. E}\ }\textbf {\bibinfo {volume} {101}},\ \bibinfo {pages} {042108} (\bibinfo {year} {2020})}\BibitemShut {NoStop}%
\bibitem [{\citenamefont {Mbeng}\ \emph {et~al.}(2024)\citenamefont {Mbeng}, \citenamefont {Russomanno},\ and\ \citenamefont {Santoro}}]{Glen_2024_scipost}%
  \BibitemOpen
  \bibfield  {author} {\bibinfo {author} {\bibfnamefont {G.~B.}\ \bibnamefont {Mbeng}}, \bibinfo {author} {\bibfnamefont {A.}~\bibnamefont {Russomanno}},\ and\ \bibinfo {author} {\bibfnamefont {G.~E.}\ \bibnamefont {Santoro}},\ }\bibfield  {title} {\bibinfo {title} {{The Quantum Ising Chain for Beginners}},\ }\href {https://doi.org/10.21468/SciPostPhysLectNotes.82} {\bibfield  {journal} {\bibinfo  {journal} {SciPost Phys. Lect. Notes}\ ,\ \bibinfo {pages} {82}} (\bibinfo {year} {2024})}\BibitemShut {NoStop}%
\end{thebibliography}

%

\clearpage{} 

\onecolumngrid 

\vspace{\columnsep} 
\begin{center}
{\large\textbf{Supplemental Material for ``Black-Hole Signatures in the Finite-Temperature Critical Ising Chain''}}{\large\par}
\par\end{center}

\begin{center}
 
\par\end{center}

\vspace{\columnsep}


\twocolumngrid

\setcounter{equation}{0}

\setcounter{figure}{0}

\setcounter{page}{1}

\setcounter{section}{0}

\global\long\def\theequation{S\arabic{equation}}%
\global\long\def\thesection{S\arabic{section}}%
\global\long\def\thefigure{S\arabic{figure}}%
%
%
%
%
%
%
%

This Supplemental Material provides technical details supporting the main text. Sec.~\ref{sec:exact} presents the exact solution of the critical transverse-field Ising chain via Jordan--Wigner and Bogoliubov transformations, including the finite-temperature retarded Green's function used in the numerical calculations. Sec.~\ref{sec:CFT} derives the low-energy continuum limit, showing that the critical spin chain reduces to a free Majorana fermion CFT with central charge $c=1/2$. Sec.~\ref{sec:geodesics} analyzes null geodesics in both pure AdS$_{3}$ and BTZ spacetimes, establishing the antipodal transport time $t_{\mathrm{trans}}=\pi$ and demonstrating that ingoing null rays inevitably fall into the BTZ horizon and do not return to the boundary. Sec.~\ref{sec:higher_curv} discusses the consistency of the fitted effective gravitational parameters $(\ell_{\mathrm{eff}},G_{\mathrm{eff}})$ with higher-curvature corrections.

\section{Exact solution}\label{sec:exact}

In this section, we present the exact solution of the transverse-field Ising chain at its critical point. We employ the Jordan--Wigner transformation to map the spin chain to a fermionic system, diagonalize the resulting Hamiltonian via a Bogoliubov transformation, and derive the finite-temperature retarded Green's function used in the numerical calculations of the main text. Without loss of generality, we consider the Hamiltonian 
\begin{equation}
H=-J\sum_{i=1}^{L}\left(Z_{i}Z_{i+1}+gX_{i}\right).
\end{equation}

\subsection{Jordan-Wigner transformation}

We use the Jordan--Wigner transformation \cite{Sachdev_1999_book_QPT,Jordan_Wigner_1928}, 
\begin{align}
X_{j} & =1-2c_{j}^{\dagger}c_{j},\\
Z_{j} & =\prod_{l<j}\left(1-2c_{l}^{\dagger}c_{l}\right)\left(c_{j}+c_{j}^{\dagger}\right),\\
iY_{j} & =\prod_{l<j}\left(1-2c_{l}^{\dagger}c_{l}\right)\left(c_{j}-c_{j}^{\dagger}\right),
\end{align}
where $c^{\dagger}$ and $c$ are fermionic creation and annihilation operators satisfying the canonical anticommutation relations. We then obtain 
\begin{align}
 & H+JgL\nonumber \\
= & -J\sum_{j}\left(-2gc_{j}^{\dagger}c_{j}+c_{j+1}c_{j}+c_{j}^{\dagger}c_{j+1}+c_{j+1}^{\dagger}c_{j}+c_{j}^{\dagger}c_{j+1}^{\dagger}\right).
\end{align}

For a system of finite size, one has to pay special attention to the fermionic boundary conditions. In particular, the original Hilbert space of the spin chain is mapped to the subspace with odd fermion number in the fermion system with periodic boundary conditions (the so-called Ramond (R) sector \cite{Francesco_CFT_Book_2012}) plus the subspace with even fermion number in the fermion system with antiperiodic boundary conditions (the Neveu--Schwarz (NS) sector \cite{Francesco_CFT_Book_2012}) \cite{Phillip_2019_scipost,Wu_2020_PRE,Glen_2024_scipost}.

\subsection{Momentum space}

We introduce the Fourier transformation 
\begin{equation}
c_{k}=\frac{1}{\sqrt{L}}\sum_{j=1}^{L}e^{-ikaj}c_{j},
\end{equation}
where the allowed momenta are fixed by the fermionic boundary condition $c_{L+1}=\pm c_{1}$ (equivalently $e^{ikaL}=\pm1$). Thus, 
\begin{align}
k & =\frac{2\pi n}{aL},\quad n\in\mathbb{Z},\quad\text{(R sector, periodic)},\\
k & =\frac{2\pi}{aL}\left(n+\frac{1}{2}\right),\quad n\in\mathbb{Z},\quad\text{(NS sector, antiperiodic)}.
\end{align}
We denote these sets as 
\begin{align}
\text{BZ}_{1}^{\text{R}} & =\left\{ \frac{2\pi n}{aL}\,\bigg|\,n\in\mathbb{Z}\right\} ,\\
\text{BZ}_{1}^{\text{NS}} & =\left\{ \frac{2\pi}{aL}\left(n+\frac{1}{2}\right)\,\bigg|\,n\in\mathbb{Z}\right\} ,
\end{align}
where in a finite system one may take any complete set of $L$ distinct momenta (e.g. $n=-\frac{L}{2},\ldots,\frac{L}{2}-1$) in each sector. Here $\text{BZ}_{1}$ denotes a choice of $L$ allowed momenta in the first Brillouin zone for the corresponding boundary condition. The inverse Fourier transform is $c_{j}=\frac{1}{\sqrt{L}}\sum_{k\in\text{BZ}_{1}^{\text{R}/\text{NS}}}e^{ikaj}c_{k}$. This convention implies the orthogonality relations $\frac{1}{L}\sum_{k\in\text{BZ}_{1}^{\text{R}/\text{NS}}}e^{ika(j-j^{\prime})}=\delta_{j,j^{\prime}}$ and $\frac{1}{L}\sum_{j}e^{i(k-k^{\prime})aj}=\delta_{k,k^{\prime}}$.

Substituting the Fourier modes into the Hamiltonian we obtain 
\begin{align}
 & H+JgL\nonumber \\
= & J\sum_{k\in\text{BZ}_{1}^{\text{R}/\text{NS}}}\Bigl[2(g-\cos ka)c_{k}^{\dagger}c_{k}+i\sin ka\bigl(c_{-k}c_{k}+c_{-k}^{\dagger}c_{k}^{\dagger}\bigr)\Bigr].
\end{align}

\subsection{Bogoliubov transformation}

We now perform a Bogoliubov transformation of the form 
\begin{equation}
\gamma_{k}=u_{k}c_{k}-iv_{k}c_{-k}^{\dagger}.
\end{equation}
We choose $u_{k}$ and $v_{k}$ such that the Hamiltonian is diagonal in terms of $\gamma_{k},\gamma_{k}^{\dagger}$ \cite{Sachdev_1999_book_QPT}, 
\begin{equation}
u_{k}=\cos\frac{\theta_{k}}{2},\quad v_{k}=\sin\frac{\theta_{k}}{2},\quad\tan\theta_{k}=\frac{\sin ka}{g-\cos ka}.
\end{equation}
At the critical point ($g=1$) we obtain 
\begin{equation}
H=-\sum_{k\in\text{BZ}_{1}^{\text{R}/\text{NS}}}\frac{\epsilon_{k}}{2}+\sum_{k\in\text{BZ}_{1}^{\text{R}/\text{NS}}}\epsilon_{k}\,\gamma_{k}^{\dagger}\gamma_{k},
\end{equation}
where $\epsilon_{k}\equiv2J\sqrt{2-2\cos ka}$.

Thus the critical transverse-field Ising chain is equivalent to the above fermion Hamiltonian in the R and NS sectors. The eigenstates and eigenvalues are 
\begin{align}
|\overrightarrow{m}\rangle_{\text{R}} & =\prod_{k\in\text{BZ}_{1}^{\text{R}}}(\gamma_{k}^{\dagger})^{m_{k}}|0\rangle,\nonumber \\
 & m_{k}=0,1,\quad\sum_{k\in\text{BZ}_{1}^{\text{R}}}m_{k}\in2\mathbb{Z}+1,\\[4pt]
E_{\text{R}}(\overrightarrow{m}) & =-\sum_{k\in\text{BZ}_{1}^{\text{R}}}\frac{\epsilon_{k}}{2}+\sum_{k\in\text{BZ}_{1}^{\text{R}}}\epsilon_{k}m_{k},
\end{align}
and 
\begin{align}
|\overrightarrow{m}\rangle_{\text{NS}} & =\prod_{k\in\text{BZ}_{1}^{\text{NS}}}(\gamma_{k}^{\dagger})^{m_{k}}|0\rangle,\nonumber \\
 & m_{k}=0,1,\quad\sum_{k\in\text{BZ}_{1}^{\text{NS}}}m_{k}\in2\mathbb{Z},\\[4pt]
E_{\text{NS}}(\overrightarrow{m}) & =-\sum_{k\in\text{BZ}_{1}^{\text{NS}}}\frac{\epsilon_{k}}{2}+\sum_{k\in\text{BZ}_{1}^{\text{NS}}}\epsilon_{k}m_{k}.
\end{align}
For numerical implementation it is convenient to use $u_{k\neq0,\pi}=\frac{\epsilon_{k}+z_{k}}{\sqrt{2\epsilon_{k}(\epsilon_{k}+z_{k})}}$ and $v_{k\neq0,\pi}=\frac{2J\sin ka}{\sqrt{2\epsilon_{k}(\epsilon_{k}+z_{k})}}$, where $z_{k}=2J(1-\cos ka)$. For $k=0,\pi$ we simply have $u_{k}=1$, $v_{k}=0$.

\subsection{Retarded Green's function}

To compute $\delta\langle n_{j}(t)\rangle$, we evaluate the two-point function 
\begin{align}
 & \langle n_{j+s}(t)n_{j}(0)\rangle\nonumber \\
= & Z^{-1}\sum_{\overrightarrow{m}}e^{-\beta E(\overrightarrow{m})}\langle\overrightarrow{m}|n_{j+s}e^{-it[H-E(\overrightarrow{m})]}n_{j}|\overrightarrow{m}\rangle.
\end{align}
Using translational invariance and time-reversal properties, one has $\langle n_{j+s}(t)n_{j}(0)\rangle=\langle n_{j}(t)n_{j-s}(0)\rangle=\langle n_{j+s}(0)n_{j}(-t)\rangle=\langle n_{j}(0)n_{j-s}(-t)\rangle$. After a lengthy but straightforward derivation we obtain 
\begin{widetext}
\begin{align}
 & G_{R}(t,s)=-i\theta(t)\langle[n_{j+s}(t),n_{j}(0)]\rangle\nonumber \\
= & -\frac{2\theta(t)}{ZL^{2}}\sum_{\overrightarrow{m}}e^{-\beta E(\overrightarrow{m})}\sum_{k,k^{\prime}\in\text{BZ}_{1}^{\text{R}/\text{NS}}}\left(u_{k^{\prime}}u_{k}-v_{k^{\prime}}v_{k}\right)^{2}\left(1-m_{k}\right)m_{k^{\prime}}\sin\left[(\epsilon_{k}-\epsilon_{k^{\prime}})t+(k^{\prime}-k)as\right]\nonumber \\
 & -\frac{2\theta(t)}{ZL^{2}}\sum_{\overrightarrow{m}}e^{-\beta E(\overrightarrow{m})}\sum_{k,k^{\prime}\in\text{BZ}_{1}^{\text{R}/\text{NS}}}v_{k^{\prime}}u_{k}\left(v_{k^{\prime}}u_{k}-v_{k}u_{k^{\prime}}\right)\left[\left(1-m_{k}\right)\left(1-m_{k^{\prime}}\right)-m_{k}m_{k^{\prime}}\right]\sin\left[(\epsilon_{k}+\epsilon_{k^{\prime}})t-(k^{\prime}+k)as\right],\label{eq:RetardedFunc}
\end{align}
where $\theta$ is the Heaviside step function. In the limit $T=0$ one has $m_{k}=0$, and this formula reduces to the result in Ref.~\cite{Bamba_2024_PRD} (there is a sign difference due to the convention that $v_{k}$ differs from that in Ref.~\cite{Bamba_2024_PRD} by an $i$ factor). 
\end{widetext}

\section{Low-energy effective description: Ising CFT}\label{sec:CFT}

In this section, we take the continuum limit of the lattice fermion Hamiltonian obtained in Sec.~\ref{sec:exact} and show that the critical transverse-field Ising chain reduces to a free massless Majorana fermion theory on a circle, which is the Ising CFT with central charge $c=1/2$. This derivation justifies the normalization convention adopted in the main text.

The dispersion relation has the low-momentum expansion 
\begin{equation}
\epsilon_{k}=2J|ka|+\mathcal{O}(k^{3}).
\end{equation}
In the low-energy subspace with this linear dispersion, we can invert the Bogoliubov transformation and write 
\begin{equation}
H\sim\sum_{k}Jka\left[2\sin\frac{ka}{2}\,c_{k}^{\dagger}c_{k}+i\cos\frac{ka}{2}\left(c_{-k}c_{k}-c_{k}^{\dagger}c_{-k}^{\dagger}\right)\right],
\end{equation}
where constant terms are omitted. We now take the continuum limit with 
\begin{align}
\psi(ai)=\frac{c_{i}}{\sqrt{a}},\quad\psi(k)=\sqrt{a}\,c_{k},\\
J\rightarrow\infty,\quad L\rightarrow\infty,\quad a\rightarrow0,\\
aL=2\pi,\quad aJ=\frac{1}{2},
\end{align}
and obtain 
\begin{equation}
H\sim aJ\int_{0}^{aL}dx\left(\psi(x)\partial_{x}\psi(x)-\psi^{\dagger}(x)\partial_{x}\psi^{\dagger}(x)\right).
\end{equation}
This Hamiltonian describes a free Majorana fermion on a circle of length $2\pi$ \cite{Francesco_CFT_Book_2012}. In particular, we recover $J=L/(4\pi)$, which justifies the normalization used in the main text.

Using the coherent-state path-integral representation, the partition function can be written as 
\begin{align}
Z & =\int\mathcal{D}(\psi^{\dagger},\psi)\,e^{-I[\psi^{\dagger},\psi]},\\
I & =\int_{0}^{\beta}d\tau\int_{0}^{2\pi}dx\,\frac{1}{2}\left(\begin{array}{cc}
\psi & \psi^{\dagger}\end{array}\right)\left(\begin{array}{cc}
-\partial_{x} & \partial_{\tau}\\
\partial_{\tau} & \partial_{x}
\end{array}\right)\left(\begin{array}{c}
\psi\\
\psi^{\dagger}
\end{array}\right).
\end{align}
The field $\psi$ is a ``complex'' Grassmann field, which we decompose into two real Grassmann fields $(\phi,\bar{\phi})$, 
\begin{equation}
\psi=\frac{\phi+i\bar{\phi}}{\sqrt{2}},\quad\psi^{\dagger}=\frac{\phi-i\bar{\phi}}{\sqrt{2}}.
\end{equation}
We then introduce two real Majorana fields $\psi$ and $\bar{\psi}$ (reusing the symbol $\psi$ for simplicity), 
\begin{equation}
\psi=\frac{\bar{\phi}-\phi}{\sqrt{2}},\quad\bar{\psi}=\frac{\bar{\phi}+\phi}{\sqrt{2}}.
\end{equation}
With complex coordinates 
\begin{equation}
z=ix+\tau,\quad\bar{z}=-ix+\tau,
\end{equation}
the action takes the standard form for a free massless real fermion, 
\begin{equation}
I[\psi,\bar{\psi}]=\int dz\,d\bar{z}\left(\psi\,\partial_{\bar{z}}\psi+\bar{\psi}\,\partial_{z}\bar{\psi}\right).
\end{equation}
This action is conformally invariant with central charge $c=1/2$ and describes the critical point of the Ising model \cite{Francesco_CFT_Book_2012}.

\begin{figure}
\includegraphics[width=1.75in]{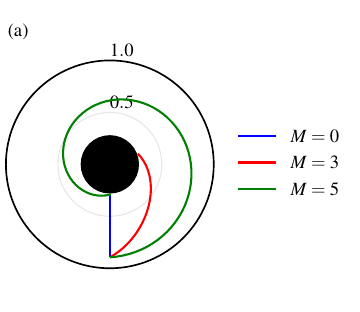}\includegraphics[width=1.75in]{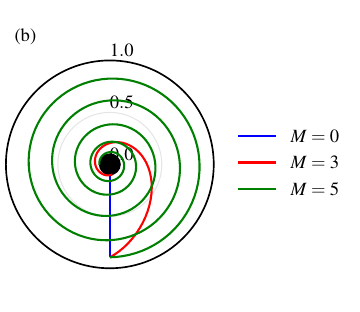}\caption{Null geodesics in BTZ spacetime. The plots show representative null trajectories in the $(t,r)$ plane, where $r\equiv\rho/\sqrt{\rho^{2}+\ell^{2}}$ is a compactified radial coordinate. A null ray sent from the boundary falls into the horizon and does not return to the boundary, illustrating why BTZ saddles do not contribute to the transported excitation. (a) and (b) correspond to horizon radii $\rho_{h}=0.3$ and $\rho_{h}=0.1$, respectively; in both cases we take $\Omega=5$, $\ell=1$, and $\rho\in(\rho_{h},2]$.}
\label{fig:geodesicBTZ}
\end{figure}

\begin{figure}
\includegraphics[width=2.9in]{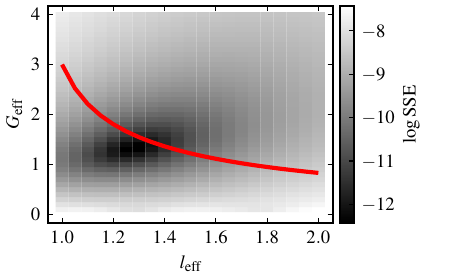}

\caption{Relation between $G_{\mathrm{eff}}$ and $\ell_{\mathrm{eff}}$ for the higher-curvature theory of Ref.~\cite{Emparan_2020_JHEP}. The red curve shows the analytic relation in Eq.~(\ref{eq:quBTZ_relation}). The color scale indicates the sum of squared errors (SSE) between the effective-theory prediction and the numerical fit to the transported excitation; darker colors correspond to smaller SSE. The curve passes through the region of minimal SSE, demonstrating consistency between the fitted $(\ell_{\mathrm{eff}},G_{\mathrm{eff}})$ and the higher-curvature correction. Here we fix $\ell_{\mathrm{classical}}=1$ and $G_{\mathrm{classical}}=3$, as in the main text.}
\label{fig:higher_curvature_relation}
\end{figure}

\section{Null geodesics in AdS$_{3}$ and BTZ spacetimes}\label{sec:geodesics}

In this section, we analyze null geodesics in the global AdS$_{3}$ and BTZ spacetimes. We first solve the null geodesic equations in pure AdS$_{3}$ and show that a null ray injected from the boundary always reaches the antipodal point at time $t=\pi$ regardless of its angular momentum, establishing the transport time $t_{\mathrm{trans}}=\pi$ used in the main text. We then analyze null geodesics in the BTZ black-hole spacetime and demonstrate that ingoing null rays inevitably fall into the horizon and do not return to the boundary. Together, these results provide the geometric basis for the claim in the main text that the BTZ saddle does not contribute to antipodal excitation transport.

\subsection{Null geodesics in pure AdS$_{3}$}\label{sec:geodesic_AdS}

We consider the global AdS$_{3}$ spacetime described by the metric 
\begin{equation}
ds^{2}=-(\ell^{2}+\rho^{2})\,dt^{2}+\frac{\ell^{2}}{\ell^{2}+\rho^{2}}\,d\rho^{2}+\rho^{2}\,d\phi^{2},\label{eq:metric_globalAdS}
\end{equation}
where $\rho\in[0,+\infty)$, $\phi\in[0,2\pi)$, and $\ell$ is the AdS curvature radius. The isometries generated by $\partial_{t}$ and $\partial_{\phi}$ yield two conserved quantities along each geodesic, 
\begin{equation}
\Omega\equiv(\ell^{2}+\rho^{2})\,\dot{t},\quad M\equiv\rho^{2}\,\dot{\phi},
\end{equation}
where a dot denotes differentiation with respect to the affine parameter $\lambda$, and $\Omega$ and $M$ are interpreted as the energy and angular momentum, respectively. Imposing the null condition $g_{\mu\nu}\dot{x}^{\mu}\dot{x}^{\nu}=0$ and eliminating $\dot{t}$ and $\dot{\phi}$ in favor of $\Omega$ and $M$, we obtain the radial equation 
\begin{equation}
\ell^{2}\dot{\rho}^{2}=\Omega^{2}-\frac{(\ell^{2}+\rho^{2})}{\rho^{2}}\,M^{2}.\label{eq:AdS_radial_eq}
\end{equation}

Introducing the rescaled affine parameter $\tilde{\lambda}\equiv(\Omega^{2}-M^{2})\lambda$, the system admits the exact solution 
\begin{align}
\rho^{2} & =\frac{\ell^{2}M^{2}+\tilde{\lambda}^{2}/\ell^{2}}{\Omega^{2}-M^{2}},\label{eq:AdS_geodesic_rho}\\
t & =\frac{\pi}{2}+\arctan\frac{\tilde{\lambda}}{\ell^{2}\Omega},\label{eq:AdS_geodesic_tau}\\
\phi & =\frac{\pi}{2}+\arctan\frac{\tilde{\lambda}}{\ell^{2}M},\label{eq:AdS_geodesic_phi}
\end{align}
where the integration constants have been chosen so that the geodesic departs from the boundary at $(t,\phi)=(0,0)$. One can verify by direct substitution that this solution satisfies both the radial equation~(\ref{eq:AdS_radial_eq}) and the conservation laws for $\Omega$ and $M$. At $\tilde{\lambda}=0$ the geodesic reaches its closest approach to the center, $\rho_{\min}=\ell|M|/\sqrt{\Omega^{2}-M^{2}}$; larger $|M/\Omega|$ corresponds to shallower penetration into the bulk.

The key property of this solution is that in the limit $\tilde{\lambda}\to+\infty$ one finds $\rho\to\infty$, $t\to\pi$, and $\phi\to\pi$, independently of the values of $M$, $\Omega$, and $\ell$. In other words, a null ray injected from the boundary always arrives at the antipodal boundary point $(t,\phi)=(\pi,\pi)$ after a coordinate time $\Delta t=\pi$, regardless of its angular momentum. This universality establishes the transport time $t_{\mathrm{trans}}=\pi$ used in the main text. By time-reversal symmetry of the AdS$_{3}$ geometry, the null ray subsequently returns to its original boundary point at $(t,\phi)=(2\pi,0)$.

\subsection{Null geodesics in BTZ spacetime}\label{sec:geodesic_BTZ}

We now consider the BTZ black-hole spacetime, whose metric takes the form 
\begin{equation}
ds^{2}=-f(\rho)\,dt^{2}+\frac{d\rho^{2}}{f(\rho)}+\rho^{2}\,d\phi^{2},
\end{equation}
with $f(\rho)=(\rho^{2}-\rho_{h}^{2})/\ell^{2}$ and horizon radius $\rho_{h}=\ell\sqrt{8GM_{\mathrm{BTZ}}}$. As in the pure AdS$_{3}$ case, the Killing vectors $\partial_{t}$ and $\partial_{\phi}$ yield the conserved energy and angular momentum, 
\begin{equation}
\Omega\equiv f(\rho)\,\dot{t},\quad M\equiv\rho^{2}\,\dot{\phi}.
\end{equation}
The null condition $g_{\mu\nu}\dot{x}^{\mu}\dot{x}^{\nu}=0$ now gives the radial equation 
\begin{equation}
\dot{\rho}^{2}=\Omega^{2}-\frac{f(\rho)}{\rho^{2}}\,M^{2}.\label{eq:BTZ_radial_eq}
\end{equation}
Eliminating the affine parameter in favor of $\rho$, the remaining geodesic equations read 
\begin{align}
\frac{dt}{d\rho} & =\pm\frac{\Omega\rho}{f\sqrt{\rho^{2}\Omega^{2}-M^{2}f}},\\
\frac{d\phi}{d\rho} & =\pm\frac{M}{\rho\sqrt{\rho^{2}\Omega^{2}-M^{2}f}},
\end{align}
where the minus sign corresponds to the ingoing branch. In contrast to the pure AdS$_{3}$ radial equation~(\ref{eq:AdS_radial_eq}), the blackening factor $f(\rho)$ vanishes at $\rho=\rho_{h}$, which qualitatively changes the geodesic structure: as illustrated in Fig.~\ref{fig:geodesicBTZ}, an ingoing null ray falls into the horizon and does not return to the boundary. This confirms that the BTZ saddle does not contribute to the antipodal transport signal. 

\section{A higher-curvature correction}\label{sec:higher_curv}

In this section, we show that the effective gravitational parameters $(\ell_{\mathrm{eff}},G_{\mathrm{eff}})$ extracted from fitting the numerical data are consistent with the higher-curvature corrections analyzed in Ref.~\cite{Emparan_2020_JHEP}.

In Ref.~\cite{Emparan_2020_JHEP}, higher-curvature corrections to BTZ black-hole spacetimes are analyzed in an effective three-dimensional description with renormalized couplings. Identifying our $(\ell_{\mathrm{eff}},G_{\mathrm{eff}})$ with the physical $\text{AdS}_{3}$ curvature radius and the corresponding renormalized Newton constant in that description, one finds from Eq.~(2.17) of Ref.~\cite{Emparan_2020_JHEP} that the backreaction parameter can be expressed in terms of $\ell_{\mathrm{eff}}/\ell_{\mathrm{classical}}$, while Eqs.~(2.40) and (2.5) express the same parameter in terms of $G_{\mathrm{eff}}/G_{\mathrm{classical}}$. Combining the two expressions yields 
\begin{equation}
\frac{G_{\mathrm{classical}}^{2}}{G_{\mathrm{eff}}^{2}}+3=4\frac{\ell_{\mathrm{eff}}^{2}}{\ell_{\mathrm{classical}}^{2}}.\label{eq:quBTZ_relation}
\end{equation}
This equation should be understood as a leading-order higher-curvature consistency relation (i.e., retaining the first nontrivial correction and neglecting higher-order terms in the curvature expansion). As shown in Fig.~\ref{fig:higher_curvature_relation}, this analytic curve passes through the region of very small sum of squared errors (SSE), indicating that the higher-curvature correction is consistent with the fitted values of $(\ell_{\mathrm{eff}},G_{\mathrm{eff}})$.
\end{document}